\documentclass[usenatbib]{mn2e}
\usepackage{color,amssymb,graphicx}
\newcommand{\be}{\begin{equation}}
\newcommand{\ee}{\end{equation}}
 
\newcommand{\xix}{ \xi_{\rm x} }

\newcommand{\zhat}{{ {\hat z} }}

\newcommand{\shat}{{\bf {\hat s} }}

\newcommand{\as}{{ a_{\rm s} }} 
\newcommand{\astwo}{{ a_{\rm s}^2 }} 
 
\newcommand{\efficiency}{{ \eta_{\rm rad} }}
 
\newcommand{\pgrav}{{ \Phi_g }} 
\newcommand{\pcent}{{ \Phi_c }} 
\def\lta{\,\raise 0.3 ex\hbox{$ < $}\kern -0.75 em
 \lower 0.7 ex\hbox{$\sim$}\,}
\def\gta{\,\raise 0.3 ex\hbox{$ > $}\kern -0.75 em
 \lower 0.7 ex\hbox{$\sim$}\,} 
\newcommand{\bratio}{\beta_*} 
\newcommand{\bc}{}
\newcommand{\cb}{}

\title[Magnetically Controlled Mass Loss]{ Magnetically controlled mass loss from extrasolar planets in close orbits} 

\author[Owen J.E. \& Adams F.C.]{James E. Owen$^{1}$\thanks{e-mail:~jowen@cita.utoronto.ca} and Fred C. Adams$^{2,3}$\thanks{e-mail:~fca@umich.edu}\\ $^1$Canadian Institute for Theoretical Astrophysics, 60 St. George Street, Toronto, Ontario, M5S3H8, Canada \\ $^2$Michigan Center for Theoretical Physics, Physics Department, University of Michigan, Ann Arbor, MI 48109, USA \\
$^3$Astronomy Department, University of Michigan, Ann Arbor, MI 48109, USA} 

\begin{document} 
\include{journals_mnras}
\pagerange{\pageref{firstpage}--\pageref{lastpage}} \pubyear{2002}

\maketitle

\label{firstpage}

\begin{abstract} 
{\bc We consider the role magnetic fields play in guiding and controlling
mass-loss via evaporative outflows from exoplanets that experience UV
irradiation. First we present analytic results that account for
planetary and stellar magnetic fields, along with mass-loss from both
the star and planet.  We then conduct series of numerical simulations
for gas giant planets, and vary the planetary field strength,
background stellar field strength, UV heating flux, and planet mass.
These simulations show that the flow is magnetically controlled for
moderate field strengths and even the highest UV fluxes, i.e.,
planetary surface fields $B_P\gta0.3$ gauss and fluxes
$F_{UV}\sim10^{6}$ erg~s$^{-1}$. We thus conclude that outflows from
{\it all} hot Jupiters with moderate surface fields are magnetically
controlled. The inclusion of magnetic fields highly suppresses outflow
from the night-side of the planet. Only the magnetic field lines near
the pole are open and allow outflow to occur. The fraction of open
field lines depends sensitively on the strength (and geometry) of the
background magnetic field from the star, along with the UV heating
rate. The net effect of the magnetic field is to suppress the mass
loss rate by (approximately) an order of magnitude. Finally, some open
field lines do not allow the flow to pass smoothly through the sonic
point; flow along these streamlines does not reach steady-state,
resulting in time-variable mass-loss. }

\end{abstract} 

\begin{keywords}magnetohydrodynamics (MHD) --- planets and satellites:
atmospheres --- planets and satellites: formation --- planets and
satellites: magnetic fields
\end{keywords}

%\newpage 
\section{Introduction} 
\label{sec:intro} 
Hot Jupiters make up an important class of extrasolar planets that
orbit their parental stars with short periods, roughly in the range
$P_{\rm orb}$ = 2 -- 6 day. They have masses comparable to Jupiter,
$M_P \sim M_J$, and display a wide range of radii and metallicity for
a given mass.  Although only about $\sim1\%$ of stars host Hot
Jupiters, these objects often transit their stars and hence their
properties -- in addition to their orbits -- can often be measured or
constrained.  Estimates have been made for their planetary radii, core
masses, and even some of their atmospheric properties.

When giant planets orbit their stars with short periods, they can be
close enough to experience substantial mass loss. The outflowing gas
can absorb UV radiation from the star and {\bc thereby increase the inferred radius of the planet's atmosphere at UV wavelengths, compared to that indicated by the actual planetary
radius (which is measured at optical wavelengths)}. This effect has
been observed in the HD209458 system (starting with \citealt{vidal}),
where current estimates indicate a mass loss rate of approximately 
$\dot{M}_P \approx 8\times10^{10}$ g/s \citep{linsky}. In addition,
the exoplanet HD189733b has been observed to experience mass loss at a
comparable rate $\dot{M}_P \sim 10^{10}$ g/s \citep{lev10}, and more
detections are expected in the near future.

The observed mass loss rates from Hot Jupiters are roughly consistent
with those expected from order of magnitude estimates. If the outflow
from the planet is controlled by the rate at which the planetary
surface gains energy from the star, the mechanical luminosity of the
outflow $GM_P\dot{M}_P/R_P$ must be balanced by the energy deposition 
rate $\efficiency F_{UV} \pi R_P^2$. Here we assume that stellar UV
radiation drives the outflow and introduce a parameter $\efficiency$
that incorporates the efficiency of energy capture and allows for the
radiation to be absorbed above the planetary surface (at $R_P$). An
order of magnitude estimate for the resulting mass outflow rate
$\dot{M}_P$ is then given by 
\begin{eqnarray}
\dot{M}_P &=& \efficiency {\pi R_P^3 F_{UV} \over G M_P}\nonumber \\ &\approx& 
10^{10} \, \, {\rm g} \, \, {\rm s}^{-1} \, \, \efficiency 
\left( {F_{UV} \over 450 \, \, {\rm erg} \, \, {\rm s}^{-1} \, \, 
{\rm cm}^{-2} } \right) \nonumber \\ &\times&
\left( {R_P \over 10^{10} \, \, {\rm cm} } \right)^{3} 
\, \, \left( {M_P \over M_J } \right)^{-1} \, , 
\label{mdotestimate} 
\end{eqnarray} where the second equality uses typical values for the planetary
properties. The fiducial UV flux $F_{UV}$ = 450 erg s$^{-1}$ cm$^{-2}$
is the flux appropriate for the quiet Sun at a distance of $a$ = 0.05
AU (Woods et al. 1998). This type of estimate has been presented
previously (for further discussion, see
\citealt{watson,lammer,barf1,barf2}; and many others).  Note that the
escape speed from the planetary surface $v_{esc}\sim50$ km/s, whereas
UV radiation generally heats gas up to temperatures $T\sim10^4$ K
(Spitzer 1978; Shu 1992) corresponding to a sound speed $\as\sim10$
km/s. Since $v_{esc}>\as$, outflows are suppressed in that the heated
gas is not free to escape, but rather must climb out of its
gravitational potential well (e.g., see the discussion of
\citealt{adams2004,owen2010,owen12} in the context of evaporation from
circumstellar disks). More sophisticated planetary outflow models have
been constructed, including chemistry, photoionization, and
recombination \citep{yelle,garcia,koskinen07,koskinen10,koskinen13},
including the effects of tidal enhancement \citep{erkaev07,mc2009},
heating from the X-rays \citep{oj12} and two-dimensional geometry
\citep{stone}.

This paper considers the problem of mass loss from planets in the
presence of magnetic fields from both the star and planet. As shown
below, magnetic fields are often expected to dominate the ram pressure
of the outflow by many orders of magnitude and cannot be neglected. On
the other hand, the effects of magnetic fields on planetary outflows
has not been well studied (previous work includes
\citealt{trammell2011,trammell2014}; and \citealt{adams2011}, hereafter
Paper I; see also \citealt{laine2008}). {\bc \citet{trammell2014} performed a set of isothermal ideal MHD simulations that included a dipole planetary field, along with rotation and the tidal field. These simulations did not include radiative transfer and the mass-loss rates were controlled by the `base-density' prescribed in the simulation domain's inner boundary. The results of these simulations followed the analytic predictions of Paper~I and semi-analytic predictions of \citet{trammell2011}. Namely, that for sufficiently strong magnetic fields outflow is confined to occur along the open field lines (from the poles) and that equatorial regions can contain a large `dead-zone' which is in magneto-static equilibrium. \citet{trammell2014} found this configuration resulted in a markedly reduced mass-loss rate compared to a pure hydrodynamic setup, where outflow can occur from the equatorial regions of the planet.

This work extends these
earlier treatments in a number of ways:  importantly we present the first multi-dimensional calculations that include EUV radiative transfer; additionally we also include more complex and
realistic geometries for the magnetic fields, extend the
parameter space under study, and by provide additional analytic
calculations to help interpret the numerical results.}

In addition to planetary outflows, however, a related body of work
exists concerning the interactions between planetary magnetospheres
and those of the stars (starting with \citealt{cuntz}). The
observational signatures of star-planet interactions include cyclic
variations of stellar activity that have the same period as the
planetary orbit; such signatures have been observed, but are often 
intermittent \citep{shkolnik2005,shkolnik2008}.

This paper is organized as follows. Section \ref{sec:parameters}
outlines the different regimes of parameter space for planetary
outflows, and defines the regime of interest here. The outflow problem
is formulated in Section \ref{sec:formulate}, along with an overview
of our numerical approach. Next we derive a collection of supporting
analytic results, including a derivation of the fraction of the
planetary surface that supports open field lines (in Section
\ref{sec:magloop}). Our main numerical results are then presented in
Section \ref{sec:results}, including the suppression of outflow on the
night side of the planet and due to lack of open field lines. Finally,
we conclude in Section \ref{sec:conclude} with a summary and
discussion of our results, along with a roadmap for further work.

\section{Partitions of Parameter Space} 
\label{sec:parameters} 

Both the star and the planet have magnetic fields (with surface
strengths $B_\ast$ and $B_P$, respectively) and outflows with mass
loss rates $\dot{M}_\ast$ and $\dot{M}_P$.  The relative strength of
these quantities determines the regime of parameter space in which the
planetary wind is launched. This section outlines the expected extent
of this parameter space. To leading order the magnetic field, on both
bodies, is taken to have a dipole form (note that we consider the
departures from this idealized case below). As a result, for purposes
of outlining the parameter space, we consider the field strength to
scale with distance according to the simple law 
\be
B = |{\bf B}| \sim B_0 \left( {R_0 \over r} \right)^{3} \, . 
\label{dipolescale} 
\ee
This form holds for both the star or the planet, where $R_0$ is the
radius of the body, $B_0$ is the surface field strength, and the
origin of the coordinate system(s) lies at its center. The mass loss
rate from either the star or the planet is constant (with radius) 
and obeys the continuity condition
\be
\dot{M} = 4 \pi r^2 \rho v \, , 
\ee
where the density $\rho(r)$ and flow speed $v(r)$ depend on the radial
coordinate.

\subsection{Dimensionless Parameters for Single Bodies}

For both the star and the planet, we can define a dimensionless
parameter $\Lambda$ that measures the ratio of ram pressure from the
outflow to the magnetic field pressure. This quantity is a function of
the radial distance $r$ from the body and can be written in the form
\be
\Lambda \equiv {2 \dot{M} v \over B^2 r^2} \, . 
\ee
The radial dependence of the magnetic field strength $B$ is given by
equation (\ref{dipolescale}), the outflow rate $\dot{M}$ is constant,
and the outflow speed $v$ is expected to be of order the sound speed
at the locations of interest. As a result, to leading order, the
parameter $\Lambda$ scales with radius according to $\Lambda\sim{r}^4$. 
To higher order, the outflow speed is a slowly increasing function of
radius and the magnetic field decreases less steeply than indicated by
equation (\ref{dipolescale}), so that $\Lambda$ increases somewhat 
more slowly than this simple scaling. 

For the star, the dimensionless parameter $\Lambda_\ast$ takes the form
\begin{eqnarray}
\Lambda_\ast &\approx& 0.004
\left( { \dot{M}_\ast \over 10^{12} {\rm g/s} } \right) 
\left( {v_\ast \over 100 {\rm km/s}} \right) 
\left( {B_\ast \over 1 {\rm G} } \right)^{-2} \nonumber\\&\times&
\left( {R_\ast \over R_\odot} \right)^{-2} 
\left( {r \over R_\ast} \right)^{4} \, ,
\end{eqnarray}
where the fiducial parameters values are chosen to be comparable to
those of the solar wind. Note that we use $v_\ast$ = 100 km/s as the
fiducial outflow speed; the asymptotic value is larger, $v_\infty \sim
400$ km/s, but the solar wind speed is smaller at the radial distances
characteristic of Hot Jupiter orbits.  The dimensionless parameter
$\Lambda_\ast$ is expected to exceed unity at a nominal radius 
$r \sim 4 R_\ast$.  Since 4-day orbits correspond to semi-major axes 
$a \sim 10 R_\ast$, the parameter $\Lambda_\ast$ will often exceed
unity at the location of the planet. However, the outflow rate from 
the star can be smaller (perhaps by a factor of 10) and the surface 
field strength can be larger (by another factor of 10), so that the 
parameter $\Lambda_\ast$ can remain less than unity out to a radius 
$r \sim 22 R_\ast$, well beyond the orbits of Hot Jupiters. As a 
result, the parameter space of interest includes both systems where
the stellar wind opens up the stellar magnetic field and systems where
the field lines remain closed (at the location of the planet).

For the planet itself, the corresponding parameter takes the form 
\begin{eqnarray}
\Lambda_P& \approx &2 \times 10^{-4} 
\left( { \dot{M}_P \over 10^{10} {\rm g/s} } \right) 
\left( {v_P \over 10 {\rm km/s}} \right) 
\left( {B_P \over 1 {\rm G} } \right)^{-2}\nonumber \\&\times& 
\left( {R_P \over 10^{10} {\rm cm}} \right)^{-2} 
\left( {r \over R_P} \right)^{4} \, ,
\end{eqnarray}
where we have used fiducial parameters appropriate for Hot Jupiters.
The dimensionless parameter $\Lambda_P$ is a function of radius
and exceeds unity at $r \sim 8.4 R_P$. As shown below, the sonic point
for the planetary wind typically falls at $r \sim 3 R_P$, so that the
parameter $\Lambda_P$ will often remain less than unity for the launch
of the outflow. The planetary field strength can be even larger,
perhaps $B_P \approx 10$ G, which would increase the crossover radius 
out to $r \sim 27 R_P$. On the other hand, for surface field strengths 
$B_P \lta 0.1$ G, the crossover radius can fall within the sonic surface. 
Since this paper focuses on magnetically controlled flow, these
calculations are only applicable for Jovian planets with surface
fields $B_P \gta 0.1$ G.

A partition of parameter space can be made by considering the four choices 
\begin{eqnarray}
&\Lambda_\ast& < 1, \Lambda_P > 1 \qquad \qquad 
\Lambda_\ast < 1, \Lambda_P < 1 \, ,\nonumber\\
&\Lambda_\ast& > 1, \Lambda_P > 1 \qquad \qquad 
\Lambda_\ast > 1, \Lambda_P < 1 \, . 
\label{quarters} 
\end{eqnarray}

For cases corresponding to the top row in equation (\ref{quarters}),
where $\Lambda_\ast < 1$, the stellar magnetic field dominates the
stellar wind at the location of the planet, and the stellar field is
essentially a dipole. In this case, the most likely configuration is
for the planet to orbit in the equatorial plane of the star, with its
pole aligned with the orbit. The magnetic field lines of the star will
be essentially vertical, in the $\zhat$ direction of the planet. One
complication that arises in this case is that the dipoles of the star
and planet can either be aligned or anti-aligned. Another complication
is that the star will not, in general, rotate with the same angular
velocity as the planetary orbit.  As a result, the field lines from
the star will tend to wrap up.

For cases corresponding to the bottom row in equation (\ref{quarters}),
where $\Lambda_\ast > 1$, the stellar wind dominates over the stellar
magnetic field at the location of the planet. The stellar wind and the
stellar magnetic field will thus be (nearly) radial at this position
(where a radial magnetic field has a split-monopole configuration).
The stellar wind and stellar magnetic field scale as $(r/R_\ast)^2$
and $R_\ast \gg R_P$, so that both are essentially constant in the
vicinity of the planet, i.e., they can be considered constant when
studying the launch of the planetary outflow.  In addition, the
angular momenta of both the spin and orbit of the planet are likely to
be (nearly) perpendicular to the equatorial plane of the star. One
likely geometry is thus for the stellar field (and wind) to point
sideways with respect to the pole of the planet.  However, many other
geometries are possible.  Another natural case to consider is where
the planet is tipped sideways so that the pole of the planet aligns
with the radial direction of the star, and hence with the direction
of both the stellar wind and stellar magnetic field. 

For cases corresponding to the right hand sides of equations
(\ref{quarters}), the planetary magnetic field is stronger than the
ram pressure of the planetary outflow. In this case, the launch of the
planetary outflow is constrained to follow the magnetic field lines,
which will (in general) be modified to include the stellar field.  If
the stellar field is stronger than the stellar wind (at the location
of the planet), the stellar field produces a nearly vertical
contribution and the launch of the wind can be described using the
formalism developed in Paper I.  If the stellar wind overwhelms the
stellar field at the planet location, then the field lines from the
planet must join onto the nearly radial (and hence nearly horizontal)
field lines from the star.

For the left hand sides of equations (\ref{quarters}), the planetary
outflows have greater ram pressure than the planetary magnetic fields.
In this case, the planetary magnetic fields become nearly radial near
the planet and the flow is nearly spherical. After leaving the
vicinity of the planet, this (nearly) spherical flow must then join
onto the environment of the star, either a dipole field that connects
to the stellar pole, or a nearly radial flow that joins onto the
stellar wind (where this latter radial flow is centered on the star).

\subsection{Dimensionless Parameters for Star-Planet Interactions} 

Next we define a collection of parameters that characterize how the
winds and magnetic fields of stars interact with the winds and
magnetic fields of the planets. For the cases where the stellar wind 
dominates over the stellar magnetic field, we must determine how the
ram pressure from the stellar wind compares to the ram pressure from
the planetary wind and to the magnetic field pressure from the
planet. The ratio of the stellar wind ram pressure, evaluated at the
location of the planet, to the ram pressure of the planetary wind is
given by
\be
\Pi_{WW} = {\dot{M}_\ast \over \dot{M}_P} 
{v_\ast \over v_P} {r^2 \over a^2} \approx 0.10 {r^2 \over R_P^2} 
\, ,  
\label{windwind} 
\ee
where $r$ is the radial coordinate centred on the planet and $a$ is
the semimajor axis of the planetary orbit. For most applications we
can take the orbit to be nearly circular, so that $a$ is also the
distance to the star. Here we expect $\dot{M}_\ast/\dot{M}_P\sim100$, 
$v_\ast/v_P \sim 10$, and $a/R_P \sim 100$, which leads to the
numerical value on the right hand side of equation (\ref{windwind}).
Using the approximate scaling law from equation (\ref{mdotestimate}),
we expect the planetary outflow rate to scale as $\dot{M}_P \sim$
$F_{UV} \sim a^{-2}$, so that the parameter $\Pi_{WW}$ should be
independent of planetary semi-major axis $a$ to leading order.  With
these fiducial values for the system properties, the planetary outflow
becomes weaker than the background outflow from the star at a radius
$r \approx 3.2 R_P$, measured from the planet, a location that falls
near the expected sonic surface.

Similarly, we find the ratio of the ram pressure from the stellar
wind, again evaluated at the location of the planet, to the pressure
provided by the planetary magnetic field. This ratio takes the form
\be
\Pi_{WB} = {2 \dot{M}_\ast v_\ast \over B_P^2 a^2} 
\left( {r \over R_P} \right)^6 \approx 3.6 \times 10^{-5} \, \, 
\left( {r \over R_P} \right)^6 \, , 
\label{windmag} 
\ee
where $r$ is the radial coordinate centered on the planet.  For the
fiducial parameter values, the ratio $\Pi_{WB}=1$ for $r\approx5.5R_P$, 
i.e., somewhat outside the expected location of the sonic surface.  
This radius (where $\Pi_{WB}$ = 1) corresponds to the magnetopause
for the planet. 

For cases where the magnetic field of the star is strong enough to
guide the stellar wind, the stellar field must be compared to both the
planetary wind and the planetary magnetic field.  The ratio of the two
magnetic fields thus provides a third dimensionless parameter that can
be written (approximately) in the form
\be
\Pi_{BB} = \left( {B_\ast \over B_P} \right)^2 
\left( {r \over R_P} \right)^6 \left( {R_\ast \over a} \right)^6 
\approx 10^{-6} \left( {r \over R_P} \right)^6 \, , 
\label{magmag} 
\ee
where $r$ is the radial coordinate centered on the planet and $a$ is
the distance to the star. The magnetic sphere of influence of the
planet thus extends out to $r \sim 10 R_P$. Note that this scaling
uses equation (\ref{dipolescale}) is thus approximate; specific 
magnetic field configurations, for both the star and planet, will 
result in modified (and non-spherical) boundaries. 

To complete the set, we define $\Pi_{BW}$ to be the ratio of the
magnetic field pressure provided by the star to the ram pressure of
the planetary wind, 
\be
\Pi_{BW} = {B_\ast^2 r^2 \over 2 \dot{M}_P v_P } 
\left( {R_\ast \over a} \right)^6 \approx 
5 \times 10^{-3} \left( {r \over R_P} \right)^2 \, , 
\label{altmagwind}
\ee 
where we have used typical values (see above) to evaluate the ratio in
the second equality. With these values, the stellar magnetic field
does not play a role within $r \sim 14 R_P$.  Keep in mind that the
the magnetic field from the star is evaluated at the location of the
planet and hence depends sensitively on the distance $a$ between the
two bodies. Since we expect $\dot{M}_P \sim a^{-2}$ (see equation
[\ref{mdotestimate}]), the parameter $\Pi_{BW}\sim a^{-4}$ (for a
dipole scaling dependence of the stellar field).  Notice also that
$\Pi_{BW}\Pi_{WB}$ = $\Pi_{BB}\Pi_{WW}$, so that the four quantities
$\Pi_{jk}$ are not independent.

Equations (\ref{windwind} -- \ref{altmagwind}) indicate that the
planetary magnetic field often tends to protect the planetary outflow,
at least until the outflow passes through the sonic surface. After
passing through the sonic point, however, the flow must join onto the
larger scale geometry that is determined by the interplay between the
stellar magnetic field and the stellar wind. In any case, it is useful
to separate the launching of the wind from its propagation at larger
distances from the planet. In particular, the launch of the wind will
often take place under conditions where the planetary magnetic fields
are strong enough to guide the flow. However, the background magnetic
field provided by the star is generally strong enough to affect the
detailed shape of the field lines and can influence the flow at the
sonic surface. In some cases, the stellar field not only changes the
location of the sonic points, but can also prevent the flow from
passing smoothly through the sonic transition (Paper I). 

\subsection{Time Dependence} 

The mass loss rates for solar-type stars are expected depend on
stellar age, so that the rate ${\dot M}_\ast$ has an approximate
time-dependence of the form 
\be
\dot{M}_\ast (t) = \dot{M}_{\ast0} 
\left( {t_w \over t_w + t} \right)^2 \, , 
\label{mdotvtime} 
\ee
where $t_w \approx 0.1$ Gyr and $\dot{M}_{\ast0}$ $\approx 2\times$
$10^{-11}$ $M_\odot$ yr$^{-1}$ \citep{wood}. Note that the starting
mass loss rate is about 2000 times the current value for the Sun. This
benchmark mass loss rate is somewhat larger than the values ($\dot{M}
\sim 10^{-13}$ $M_\odot$ yr$^{-1}$) considered ``typical'' for
weak-lined T Tauri stars (e.g., \citealt{gunemer}). On the other hand,
this initial mass loss rate $\dot{M}_{\ast0}$ is somewhat smaller than
the rates expected for classical T Tauri stars; these object exhibit a
wide range of values $\dot{M}\sim10^{-8}-10^{-10}$ $M_\odot$ yr$^{-1}$
(e.g., \citealt{hart}). We thus expect equation (\ref{mdotvtime}) to
provide a good estimate for the average mass loss rates as a function
of time, but the variance will be large for early times (especially
the T Tauri phases).

The magnetic field strength is observed to scale (roughly) with the
stellar rotation rate, and both decrease with time.  One version of
this scaling law is a magnetic Bode's law \citep{bali}, which shows
that the stellar magnetic moment scales with the stellar angular
momentum, so that we expect a scaling law of the general form 
\be
B_\ast \sim \Omega_\ast^{1/2} \qquad {\rm where} \qquad 
\Omega_\ast \sim t^{-1/2} \, , 
\ee
where the second expression is the well-known relationship for
spin-down of stars \citep{skum}. Taken together, these two results
indicate that $B_\ast \sim t^{-a}$, where the index $a \approx 1/4$. 
Since the stellar mass loss rates scales as $\dot{M}_\ast\sim$
$t^{-2}$, and $B_\ast^2 \sim t^{-1/2}$, the dimensionless parameter
$\Lambda_\ast \sim t^{-3/2}$.  This result would indicate that
$\Lambda_\ast$ would be larger in the past. However, this scaling only
applies to relatively old stars, i.e., these results cannot be
extrapolated back to the early pre-main-sequence phases.

In contrast, T Tauri stars, with ages of a few Myr, have surface
fields $B_\ast \sim 2500$ G \citep{jk2009}.  These young stars often
have substantial components of their magnetic field in higher order
multipoles \citep{gregory2010,gregory2011}, whereas only the dipole
component is relevant at the location of the planets.  Nonetheless,
the dipole component is still expected to have a large field strength
of $B\sim1000$ G. Compared to Solar values, T Tauri stars thus have
values of $B_\ast^2$ that are larger by a factor of $\sim10^6$, but
the outflow rates are larger, on average, by only a factor of
$\sim2000$. This scaling would indicate that the parameter
$\Lambda_\ast$ is smaller for young stars by a factor of $\sim1000$.
But weak-line T Tauri stars have outflow rates that are weaker than
this average value, and hence have even smaller values of
$\Lambda_\ast$. Classical T Tauri stars can produce much larger
outflow rates, more than $10^6$ times the current Solar value, and
could thus have smaller values of $\Lambda_\ast$.

Taken together, the above results indicate that the values of the
dimensionless parameters $(\Lambda,\Pi)$ are likely to vary
substantially with the age of the star/planet system. Further, these
parameters can be either larger or smaller in the past, and are
expected to vary from system to system.

\section{Formulation of the Outflow Problem} 
\label{sec:formulate} 

For the sake of definiteness, we consider a simple magnetic field
configuration consisting of two components. The planet has a dipole
field with strength $B_P$.  In addition, the stellar field at the
location of the planet has a contribution that we model as a constant
field that points along the pole of the planet, i.e., 
\be
{\bf B} = \bratio B_P \zhat \qquad {\rm where} \qquad 
\bratio \equiv {B_\ast \over B_P} \left( {R_\ast \over a} \right)^3\,,
\label{betadef} 
\ee
where $B_P$ and $B_\ast$ are the surface field strengths on the planet
and the star, respectively. With this choice of stellar field
component we have restricted ourselves to cases where
$\Lambda_\ast<1$. Additionally, we assume that this stellar field
structure protects the planet from the stellar wind, again restricting
ourselves to the region of parameter space where $\Pi_{WW}\ll 1$. 
Thus, we do not include a stellar wind component in our detailed
calculations. Therefore, the aim of this initial study is to
investigate the interplay between the magnetic field strength and
structure and the evaporative flow (i.e., varying $\Lambda_P$ and
$\Pi_{BW}$). Specifically, we aim to understand under what conditions
the launching of the evaporative flow is controlled by the magnetic
field ($\Lambda_P \ll 1$, $\Pi_{BW}\ll1$) or the evaporative flow is
strong enough to fully disrupt the magnetic field structure allowing
quasi-spherical outflow ($\Lambda_P\gg 1$, $\Pi_{BW}\gg 1$).  For 
cases where the magnetic field controls the flow geometry, planetary 
mass-loss can be significantly suppressed relative to the
quasi-spherical outflows (see Paper I) that are commonly used in
modelling planetary evaporation
\citep[e.g.][]{lammer,koskinen07,mc2009,oj12}.

%The main features of this calculation include the following: 
%
%One of the main characteristic of this work is that the flow is
%magnetically controlled. The MHD code used herein accounts for the
%back reaction of the fluid on the field lines, i.e., the magnetic
%field configuration is allowed to evolve in response to the outflow.
%In practice, however, the field evolution is negligible, so that the
%magnetic field configuration does not change. We note that this trend 
%remains true even for the more extreme UV fluxes. 
%
%The outflow is dominated by the field lines, and hence the
%streamlines, that originate from the day side of the planet. The
%presence of the magnetic fields prevent the zonal flows of the planet
%from carrying heat to the night side, so that only one half of the
%planet supports outflow. This trend is illustrated by Figure X, which
%shows outflow as a function of planetary longitude. Since only the day
%side of the planet contributes, as shown in the figure, we consider
%only the day side of the planet for the remainder of this paper. 
%
%The thermodynamics of the outflow is modeled using the following 
%prescription: 
%
%Radiative transfer in the simulations is modeled as follows: 
%(for further detail, see Appendix \ref{sec:radtransfer}). 
%
%The basic elements of our numerical integration scheme are described 
%in Appendix \ref{sec:numscheme}. 

Even with our restricted choice of interest, the parameter space for
planetary outflows is large. We must specify the planet properties,
including the planetary mass $M_P$, radius $R_P$, and surface field
strength $B_P$. We must also specify the stellar properties that
define the environment that the planet resides within, i.e., the
background stellar field strength (determined by the parameter
$\bratio$) and the stellar UV flux $F_{UV}$ evaluated at the location of
the planet. For most of this work, we focus on Hot Jupiters with mass
$M_P = 1.0 M_J$ and $R_P = 10^{10}$ cm, although we vary the planet
mass for one series of simulations.  Note that this value for the
radius is somewhat larger than that of Jupiter itself, where this
radius anomaly is well known for Hot Jupiters
\citep{boden2003,laugh2011}. With these choices, the relevant
parameter space is given by $(B_P, \bratio, F_{UV})$.

In this work we consider both numerical calculations and analytic
studies, with a focus on the latter.  We use supporting analytic
calculations as a guide to explore the underlying physics and
interpret the results of the numerical calculations, as well as to
draw inferences out of the range of our simulations. As a result, for
our analytic work we follow Paper I and consider the magnetic field to
be static and force-free (and this approximation is largely vindicated
by the numerical simulations).  In our numerical calculations we do not evolve the full energy equation and instead use a simplified thermal update \citep{ivine}. This simplification restricts our initial calculations presented within to the radiative-recombination regime \citep{mc2009} and as such high UV fluxes ($\gtrsim 10^{5}$ erg s$^{-1}$). Additionally evolving the energy equation in multi-dimensions with ionizing chemistry and radiative transfer is computationally challenging and will be done in our next study. 

\subsection{Numerical Calculations}\label{sec:numerical_setup}

In the numerical studies we solve the Radiation-MHD problem in the
ideal MHD limit (i.e., the magnetic structure is allowed to respond 
to the flow). In addition to the standard ideal-MHD equations, we
additionally evolve the ionization fraction in the flow and the
radiative transfer problem for ionizing photons. The time evolution 
of the ionization fraction is given by
\begin{equation}
\frac{{\rm D}X}{{\rm D}t}=(1-X)(\Gamma +n_eC)-
Xn_e\alpha_r\,,
\label{eqn:Ion_balance}
\end{equation}
where $X$ is the ionization fraction, $n_e$ is the electron density,
$\Gamma$ is the photoionization rate, $C$ is the collisional
ionization rate, and $\alpha_r$ is the recombination rate. We
calculate the photoionization rate $\Gamma$ assuming a monochromatic
spectrum with a frequency of $h\nu_{13.6}=13.6$~eV, such that
\begin{equation}
\Gamma=\frac{F_{UV}}{h\nu_{13.6}}\sigma_{13.6}\exp(-\tau)\,,
\end{equation}
where $\sigma_{13.6}$ is the photoionization cross section at energy 
$h\nu_{13.6}=13.6$eV \citep{osterbrock}. Additionally, $\tau$ is the
optical depth to ionizing photons and is defined according to 
\begin{equation}
\tau=\sigma_{13.6}N_{HI}\,,
\end{equation}
where $N_{HI}$ is the neutral Hydrogen column density.

{\bc Our choice of a monochromatic spectrum at 13.6eV is chosen for numerical convenience. However, since our parameter range of interest is in the recombination balance regime the chosen photon energy does not control the level of heating/ionization in our setup. It is merely the number (rather than their energy) of ionizing photons that controls the level of ionization and hence the mass-loss rates. This approximation obviously cannot be extended to arbitrarily 
low UV fluxes (but such low fluxes are not considered here). \citet{mc2009} choose a characteristic energy of 20eV. Thus, there is a small correction factor in the number of ionizing photons of $1.47$ when comparing fluxes in terms of energy per-unit time and a small difference in the mass-loss rates (which goes approximately as the square-root of the number of ionizing photons) of 1.2, much smaller than the differences we find due to the presence of the magnetic fields.}

Our numerical calculations are performed using a modified version of
the {\sc zeus-MP} MHD code \citep{stone_hd,stone_mhd,hayes06}, where
we additionally solve equation~(\ref{eqn:Ion_balance}), along with the
radiative transfer of ionizing EUV photons. Our radiative transfer
scheme is detailed in Appendix~\ref{sec:radtransfer} and our numerical
approach is described in detail in Appendix~\ref{sec:numerical}.
Essentially, we assume that the recombination time is short compared
to the flow time and the ionizing photons have a mean-free path that
is short compared to the flow length-scale at the ionization front. As
such, this set of assumptions restricts us to the largest UV fluxes
($\gtrsim~10^{5}$~erg~s$^{-1}$), where the gas is close to
radiative-recombination equilibrium; however, it allows us to simplify
the thermal structure in the flow where ionized gas is assumed to be
isothermal at $10^{4}$~K and neutral gas is taken to be isothermal at
$10^{3}$~K. We stress that this restriction in EUV flux does not
prevent us achieving our goal for this work: Any outflow that is
magnetically dominated at high fluxes, will also be magnetically
dominated at lower fluxes since the mass-loss rate increases with
increasing flux.

{\cb All of the numerical calculations are performed on a 2D spherical grid
(${r,\theta}$). Note that for the
simulation where we include both the day and night side of the planet, the use of a
2D grid involves a greater degree of approximation and cannot be considered globally axisymmetric (see Section~5.1). We emphasise that planetary evaporation is a fundamentally 3D process. In particular, rotation cannot be included in our simulations with both a day and night side as this would violate the `pseudo-symmetry' of our setup. Furthermore, the centrifugal force cannot be included at all in such a simulation; however, it is expected to be very small in planetary evaporation \citep{stone,mc2009,oj12}.} {\bc In all cases we take the symmetry axis of the planetary dipole to be perpendicular to the orbital plane. Additionally any contribution from the stellar magnetic field $\beta_*>0$ is also assumed to be perpendicular to the orbital plane (see Equation~\ref{betadef}). In our plots we adopt the standard Cartesian to spherical co-ordinate system mapping, with the z-axis taken to be the symmetry axis of the dipole and the star is located along the positive x-axis.} Our computational domain has an inner
boundary at $10^{10}$~cm and at outer boundary at $1.5\times10^{11}$
cm. For reference, note that the sonic radius for an isothermal
($10^{4}$~K), spherical flow is $\sim 3\times10^{10}$~cm for a Jupiter
mass planet. The radial grid is non-uniform and is of size $N_r=128$,
where the resolution at the inner boundary has approximately $100$~km
sized cells, sufficient to resolve the scale height of the underlying
bolemetrically heated atmospheres. In the angular direction we use a
uniform grid with 64 cells per quadrant. At the inner boundary we
apply fixed boundary conditions where the density is set to
$10^{-11}$~g~cm$^{-3}$, the temperature is set to $10^{3}$~K, magnetic
field is set to a dipole of strength $B_P$, and the ionization
fraction is set to $X=10^{-5}$. On the outer boundary we adopt outflow
boundary conditions, but include the contribution from the background
stellar field if $\bratio>0$. Finally, on the angular boundaries we
adopt the appropriate symmetry boundary conditions. In order to
isolate the effects of the magnetic field, we neglect the small
contributions from planetary rotation and the stellar gravitational
field.

We initialise the simulations to be isothermal at $10^{3}$~K, with a
hydrostatic density structure close to the planet. {\bc At larger radii we
fill the grid with a low density gas (which is optically thin to 13.6
eV photons) that falls off with density as $\rho \sim r^{-2}$; this
density profile is normalized such that the plasma beta in the grid is
larger than $10^{-4}$ in order to prevent very short numerical
time-steps. This radius where we transfer from the hydrostatic density structure to the power-law fall off depends on the initial magnetic field strength, but typically occurs around $\sim 2$ planetary radii. }Note that, in general, our simulations evolve towards
steady-state solutions, so that this initial density structure is
purely a matter of convenience. We then evolve the flow system for
$\sim$15 flow crossing times. Unless specifically stated otherwise,
all of the results from the simulations described herein are measured
after 13 flow crossing times. In general, the flow reaches
steady-state after only 2 -- 3 flow crossing times.

\section{Analytic Expectations: Magnetic Loops and Open Fields Lines} 
\label{sec:magloop} 

This section calculates the hydrostatic structure of coronal plasma
following magnetic loops on planetary surfaces, we can then use this
result to understand under what limits the flow will be controlled by
planet's dipole. The formulation is general, but the application is
made for Hot Jupiters. We start by considering dipole magnetic field
configurations, but the results can be generalized to include
quadrupole, octupole, and more general cases (although this approach
is limited to cases with azimuthal symmetry).  More specifically, we
find analytic expressions for the shape of the magnetic field lines,
the coordinates following the field lines, and the pressure integrated
along the field lines.  These results are then used to determine the
fraction of the surface that supports closed field lines, the radial
extent of the loops, and the corresponding volume of the trapped
magnetic region.

\subsection{Basic Formulation} 
\label{sec:basicloop} 

For a rotating system, the effective gravity ${\bf g}$ is given by 
\begin{eqnarray}
{\bf g}& =& (g_r, g_\theta, g_\phi)\nonumber\\& = &
\left( - {GM_P \over r^2} + \Omega^2 r \sin^2\theta, 
\Omega^2 r \sin\theta \cos\theta, 0 \right) \, , 
\end{eqnarray}
where $\Omega$ is the planetary rotation rate, and where the rotation 
axis coincides with the $\zhat$ direction of the coordinate system. 
Note that this analytic treatment includes rotational effects
($\Omega\ne0$) so that we can assess their importance. Since
rotational effects are small (see below), in our numerical treatment
(see the following section) we neglect rotation in order to isolate
the effects of the magnetic fields.

We assume that the plasma is isothermal with sound speed $\as$. As
discussed above the temperature is expected to be about $T_C \sim 10^4$ K 
and the magnetic field strengths are typically in the range
$B=1 - 10$ G.  The sound speed is thus $\as \sim 10$ km s$^{-1}$.

If we assume that the coronal plasma is in hydrostatic equilibrium, 
the pressure along a given magnetic loop takes the form 
\be
P(s) = P_0 \exp \left[ {1 \over \astwo} 
\int_0^s {\bf g} \cdot \shat \, ds \right] \,  ,
\ee
where the integral starts at the planetary surface and 
continues to the point $s$ along the magnetic loop. 
One can show that 
\be
{\bf g} \cdot \shat \, ds = {1 \over B} 
{\bf g} \cdot {\bf B} \, ds = g_r dr + 
{B_\theta \over B_r} g_\theta dr \, .
\ee
The pressure integral then becomes 
\be
P(s) = P_0 \exp \left[ {1 \over \astwo} \left( 
\int_{R_P}^r g_r dr \, + \,  
\int_{R_P}^r {B_\theta \over B_r} g_\theta dr \right) \right] \, . 
\label{pressure} 
\ee
In order to determine the pressure, we must evaluate the integrals 
\be 
I_1 = {1 \over \astwo} \int_{R_P}^r 
\left( - {GM_P \over r^2} \right) dr \, = 
{G M_P \over \astwo R_P} \left( {R_P \over r} - 1 \right) \, , 
\ee
\be
I_2 = {1 \over \astwo} \int_{R_P}^r \Omega^2 r \sin^2\theta dr\,,
\ee
and
\be
I_3 = {1 \over \astwo} \int_{R_P}^r 
{B_\theta \over B_r} \Omega^2 r \sin\theta \cos\theta dr \, . 
\ee
Next we define dimensionless quantities 
\be 
\pgrav \equiv {G M_P \over R_P \astwo} \, , \;\;\;
\pcent \equiv {1 \over 2} \left( {\Omega R_P \over \as} \right)^2 \, , 
\;\;\; {\rm and} \;\;\; \xi \equiv {r \over R_P} \, . 
\label{nddefs} 
\ee
If we take typical parameters so that $M_P$ = 1.0 $M_J$, 
$R_P = 10^{10}$ cm, $\as$ = 10 km s$^{-1}$, and period $P_{rot}$
= 4 days, then $\pgrav \sim 13$ and $\pcent \sim 0.017$. The 
parameters of this problem thus obey the ordering 
\be
\pcent \ll 1 \ll \pgrav \, . 
\ee
We next note that 
\be 
I_1 = \pgrav \left( {1 \over \xi} - 1 \right) \, , 
\ee 
so that $I_1$ is the same for all magnetic field configurations. 
The remaining two integrals have the form 
\be
I_2 = \pcent J_2 \quad {\rm where} \quad 
J_2 = 2 \int_1^\xi \sin^2\theta \, \xi d\xi \,,
\ee
and 
\be
I_3 = \pcent J_3 \quad {\rm where} \quad 
J_3 = 2 \int_1^\xi {B_\theta \over B_r} 
\sin\theta \cos\theta \, \xi d\xi \, . 
\ee
We thus need to evaluate $J_2$ and $J_3$ for a given form of the
magnetic field configuration. Note that along each field line, the
angle $\theta$ depends on the dimensionless radius $\xi$, as
determined by the field geometry. Once all of the dimensionless
integrals have been evaluated, the pressure is then given by 
\be
P(s) = P_0 \exp \left[ \pgrav \left( {1 \over \xi} - 1 \right) + 
\pcent \left( J_2 + J_3 \right) \right] \, . 
\label{pressurend} 
\ee
As we show below for dipole field configurations, the dimensionless
integrals $J_2$ and $J_3$ combine to take the form 
\be
J_2 + J_3 = x^2 - x_P^2 \, , 
\ee
where $x$ = $\xi\sin\theta$ is evaluated at the field point
$(\xi,\theta)$ and where $x_P$ is the coordinate at the planetary
surface that connects to the field point along a magnetic field line. 

\subsection{Dipole Field Configurations} 
\label{sec:dipole} 

For the case of dipole fields, the magnetic field components 
have the form 
\be
B_r = B_0 \xi^{-3} 2 \cos \theta 
\quad {\rm and} \quad 
B_\theta = B_0 \xi^{-3} \sin \theta 
\label{bdipole} 
\ee
where we have defined $\xi = r/R_P$. The magnetic field lines 
follow lines of constant values of the coordinate `$q$' (e.g., 
\citealt{ag2012,adams2011}), so that 
\be 
q = \xi^{-1} \sin^2 \theta = \sin^2 \theta_0 = constant \, . 
\ee 
The constant $q$ is thus determined by the polar angle ($\theta_0$) of 
the loop at the planetary surface (the location of the footpoint). 
With these specifications, the integral $J_2$ becomes 
\be
J_2 = 2 \int_1^\xi \sin^2\theta \xi d\xi = 2 q 
\int_1^\xi \xi^2 d\xi = {2 \over 3} q \left( \xi^3 - 1 \right) \,.
\ee 
Similarly, the integral $J_3$ becomes 
\begin{eqnarray}
J_3 &=& 2 \int_1^\xi {\sin\theta \over 2 \cos\theta} 
\sin\theta \cos\theta \xi d\xi \, =
\int_1^\xi \sin^2\theta \xi d\xi \,\nonumber \\  &=& 
q \int_1^\xi \xi^2 d\xi = {q \over 3} \left( \xi^3 - 1 \right) \, . 
\end{eqnarray}
As a result, the sum of the two integral simplifies to the form 
\be
J_2 + J_3 = q \left( \xi^3 - 1 \right) = x^2 - x_P^2 \, , 
\ee 
where $x_P$ is the value at the planetary surface. The 
pressure can then be written 
\be
P(\xi) = P_0 \exp \left[ \pgrav \left( {1 \over \xi} - 1 \right) 
+ \pcent \, q \left( \xi^3 - 1 \right) \right] \, .  
\label{pressuredip} 
\ee

\begin{figure} 
%\figurenum{1}
\centering
\includegraphics[width=\columnwidth]{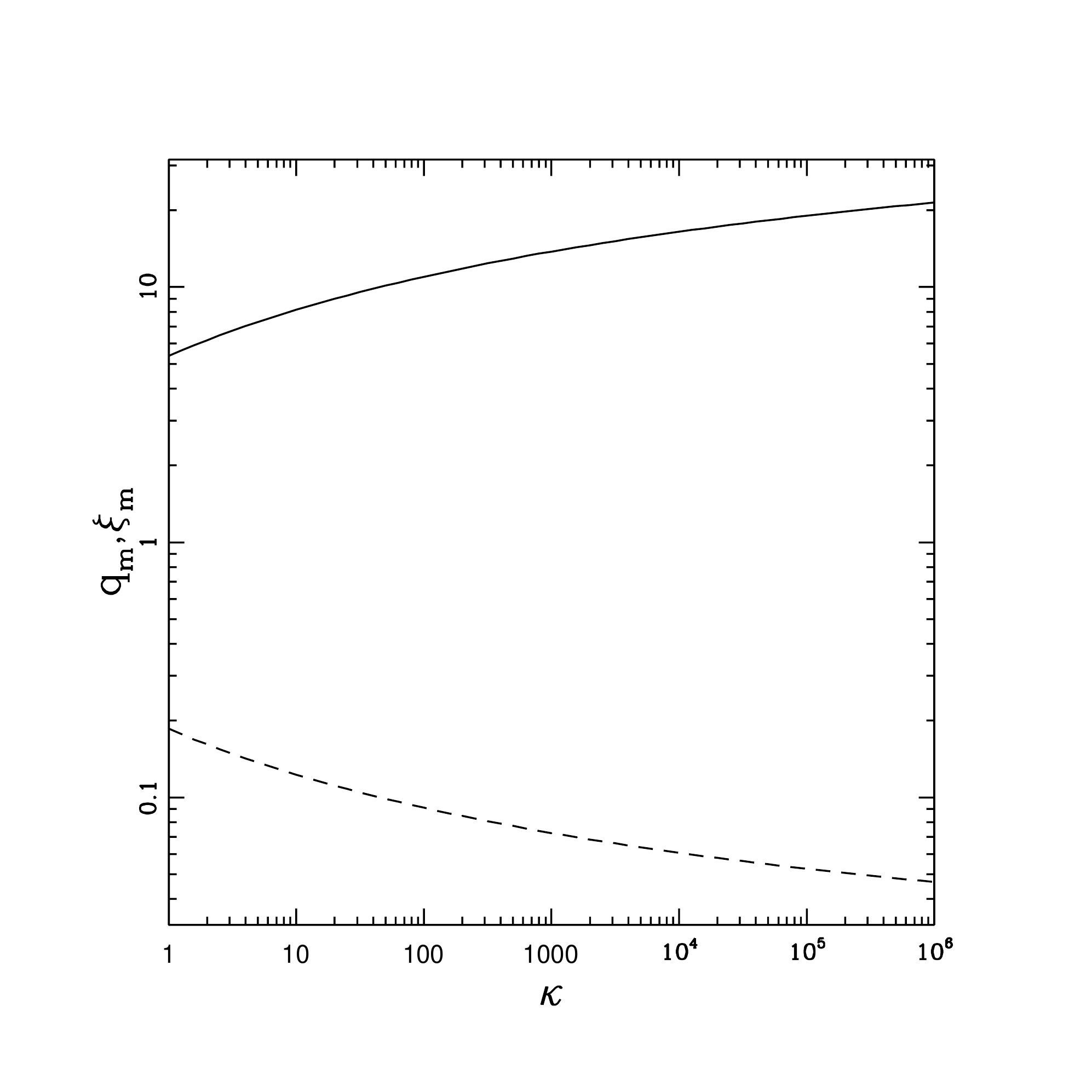}
\caption{Coordinate $q_m$ of the critical magnetic field line 
(dashed curve) and the corresponding maximum radial extent $\xi_m$ 
of the flux loop (solid curve).  Both quantities are plotted versus
the parameter $\kappa$ = $B_0^2/(8 \pi P_0)$, which measures the 
relative strength of the magnetic field at the planetary surface.
The potentials have fixed values $\pgrav$ = 13 and $\pcent$ = 0.017. }
\label{fig:loopsize} 
\end{figure}

The planetary surface will support both closed field lines and open
field lines, with a critical magnetic field line delineating the
boundary between them.  We can set the value of the critical
streamline, labeled by $q_m$, by requiring that the magnetic pressure
is greater than the gas pressure (from equation [\ref{pressuredip}])
at all points along the critical field line. At the point of equality, 
\begin{eqnarray}
&& P_0 \exp \left[  \pgrav  \left( {1 \over \xi} - 1 \right) 
+ \pcent \, q \left( \xi^3 - 1 \right) \right] = \nonumber \\ &&  \quad {B_0^2 \over 8 \pi} \left( 4 \cos^2 \theta + \sin^2 \theta \right) \xi^{-6} \, . 
\end{eqnarray}
In general, the magnetic field pressure decreases faster than the gas
pressure, so we want to evaluate the above expression at the largest
radius $\xi$ of the magnetic loop. For dipole fields, considered here,
the largest value of the radius is given by $\xi=1/q$, which occurs
where $\sin\theta=1$ and $\cos\theta=0$ (along the equator). The above
equation becomes 
\be
\pgrav \left( q - 1 \right) + \pcent \left( q^{-2} - q \right) = 
\log \left[ {B_0^2 \over 8 \pi P_0} \right] + 6 \log q \, . 
\label{trans} 
\ee
One must solve the transendental equation (\ref{trans}) to find the
critical value of the variable $q=q_m$ that labels the critical
magnetic field line. The result depends on the values of $\pgrav$
and $\pcent$, as defined above, as well as the ratio $\kappa$ of
the magnetic field pressure to the gas pressure at the planetary surface
(the base of the magnetic loop). Specifically we define 
\be
\kappa \equiv {B_0^2 \over 8\pi P_0} \, . 
\ee
For fixed values of $\pgrav$ = 13 and $\pcent$ = 0.017, Figure
\ref{fig:loopsize} shows the maximum radial extent of the loops as a
function of the parameter $\kappa$. If we take the limit $\pcent\to0$ 
and then consider $q \ll 1$, equation (\ref{trans}) can be solved for
the critical value of the coordinate, i.e., 
\be
q_m \approx \kappa^{-1/6} \exp\left[-\pgrav \right]\,.
\ee

The critical value of the coordinate $q_m$ corresponds to a critical 
value of the polar angle $\theta_m$ on the planetary surface, i.e., 
\be
q_m = \sin^2 \theta_m \,.
\ee 
The fraction $F_{AP}$ of the planetary surface that supports open field 
fields is given by 
\begin{eqnarray}
F_{AP} &=& 1 - \cos\theta_m = 1 - \left( 1 - \sin^2\theta_m \right)^{1/2} 
\nonumber \\ &=& 1 - \left(1 - q_m \right)^{1/2}\,, 
\end{eqnarray}
which reduces to the approximate form 
\be
F_{AP} \approx {1 \over 2} \kappa^{-1/6} \exp \left[ -\pgrav / 6 \right] \,.
\label{fractherm} 
\ee
This fraction $F_P$ corresponds to the the open field lines produced 
due to the hot plasma opening up the magnetic field, which has a purely 
dipole form. For the case where the field also has a (straight) background 
component (e.g., due to the star), a fraction of the planetary surface will 
support open field lines even in the limit of zero temperature. This fraction 
$F_{BP}$ is given by (see Paper I) 
\be
F_{BP} = 1 - \left[ 1 - {3 \bratio^{1/3} \over 2 + \bratio} \right]^{1/2} 
\approx {3 \over 4} \bratio^{1/3} \,,
\label{fracfield} 
\ee 
where the second equality assumes $\bratio \ll 1$. 

By comparing equations (\ref{fractherm}) and (\ref{fracfield}), we can
determine which process is dominant in producing open field lines,
thermal opening of magnetic loops or the underlying field geometry
(including the stellar background field). Since $\pgrav \sim 10-12$,
the fraction $F_{AP} \sim 0.1 \kappa^{-1/6}$, whereas the fraction
$F_{BP} \sim 0.1$ (since the parameter $\bratio \sim 0.001$). In most
cases, more field lines are open due to the background stellar field
than are opened up by the plasma pressure. However, the latter effect
scales as $\kappa^{1/6}$ so that sufficiently hot plasma temperatures 
can also lead to additional open field lines.  In general, the ratio 
of the two areas is given by 
\be
{F_{BP} \over F_{AP}} = {3 \over 2} \exp\left[\pgrav/6\right] 
\bratio^{1/3} \kappa^{1/6} \approx 11\bratio^{1/3} \kappa^{1/6}\,.
\ee
Next we note that $\kappa \sim B_P^2$ and $\bratio \sim B_\ast/B_P$, 
where $B_P$ is the surface field strength on the planet, so that 
the ratio is independent of the planetary field strength. However, 
this expression is only valid in the regime where the planetary 
field strength is large enough to control the flow; in practice, 
one needs $B_P \gta 0.3$ gauss for the largest expected stellar 
UV fluxes. 

Finally, we can estimate $\kappa$ in terms of the incident UV
flux. Ignoring advection in equation~(\ref{eqn:Ion_balance}) and
negltecting collisional ionization (which is known to be sub-dominant
-- see \citealt{mc2009}). Then equation~(\ref{eqn:Ion_balance}) can
simply be expressed by balancing the number of incoming photons with
the number of recombinations such that:
\begin{equation}
\frac{F_{UV}}{h\nu_{13.6}}=\int_0^{\infty} n^2\alpha_r d\ell
\end{equation}
where $\ell$ is a ray extending from the star to the planet. If we
consider the ray reaching the sub-stellar point of the planet, and
drop the contribution from the $\Phi_c$ term, then equation 
(\ref{pressuredip}) can be used to express the number density
$n$ in the form 
\begin{equation}
n(\xi)=\left(\frac{P_0}{\mu_{\rm mmw} a_s^2}\right)
\exp\left[\Phi_g\left(\frac{1}{\xi}-1\right)\right]\,,
\label{eqn:ion_balance1}
\end{equation}
{\bc where $\mu_{\rm mmw}$ is the mean molecular weight ($\mu_{\rm mmw}=0.5m_h$ for a gas 
consisting of pure ionized hydrogen gas)}. Thus, for the sub-stellar 
point, equation~(\ref{eqn:ion_balance1}) may be written as 
\begin{equation}
\frac{F_{UV}}{h\nu_{13.6}}=\alpha_r
\left(\frac{P_0}{\mu_{\rm mmw} a_s^2}\right)^2R_P
\int_1^\infty\exp\left[2\Phi_g\left(\frac{1}{\xi}-1\right)\right]d\xi\,.
\label{eqn:ion_balance2}
\end{equation}
The integral in equation~(\ref{eqn:ion_balance2}) formally diverges 
due to the finite pressure at infinity, which arises because of the 
hydrostatic assumption. Such a finite pressure is not physical
\citep[e.g.,][]{parkerspiral} and the recombinations are expected to be
dominant close to the planet in a realistic scenario. As a result, after
truncating the integral after several scale heights, one finds that 
\begin{equation}
\frac{F_{UV}}{h\nu_{13.6}}\approx\frac{\alpha_r}{2\Phi_g}
\left(\frac{P_0}{ \mu_{\rm mmw} a_s^2}\right)^2R_P\,,
\label{eqn:density_if}
\end{equation}
where the result is independent of the truncation point. 
We can then cast $\kappa$ in terms of the flux as
\begin{eqnarray}
\kappa&\approx&50\left(\frac{B_0}{1\mbox{~gauss}}\right)^2
\left(\frac{F_{UV}}{10^{4}\mbox{ erg s}^{-1}}\right)^{-1/2}\nonumber \\ &\times &
\left(\frac{\Phi_g}{13}\right)^{-1/2}
\left(\frac{R_P}{10^{10}\mbox{ cm}}\right)^{1/2}\,.
\end{eqnarray}
Therefore, we expect $\kappa\gg1$ for Hot Jupiters with moderate
magnetic field strengths.

\subsection{Volume of Loop Regions} 
\label{sec:volume} 

The volume of the regions that support closed magnetic loops is another 
interesting quantity in this problem. For dipole field configurations, 
the volume of the loop region is given by an integral of the form
\be
V = 4\pi \int_0^{\mu_m} d\mu \, \int_1^{\xix} \xi^2 d\xi \, ,
\label{volumedef} 
\ee
where we have assumed azimuthal symmetry {\bc and $\mu=\cos\theta$}.  For the case of dipole
fields, the loop reaches its point of maximum extent at the equator
where $\theta$ = $\pi/2$, and we have used the fact that the loops are
symmetric with respect to the equatorial plane.

The largest magnetic loop that remains closed defines the outer
boundary of the loop region.  This loop intersects the planetary
surface at $\xi$ = 1 for polar angles given by $\mu_m$, where $\mu_m$
= $\cos\theta_m$, and reaches its point of maximum extent at
dimensionless radius $\xi_m$.  These defining quantities ($\mu_m$,
$\xi_m$) depend on the parameters of the problem as shown above. Note
that the upper end $\xix$ of the radial integration in equation
(\ref{volumedef}) is given by the intersection of a ray (determined by
the angular variable $\mu$) with the outermost magnetic loop. Note
that in general $\xi_m\ne\xix$.

The first integral can be immediately evaluated to obtain the form 
\be
V = {4\pi \over 3} \int_{0}^{\mu_m} d\mu \, 
\left( \xix^3 - 1 \right) \, . 
\label{volumeone} 
\ee

The field line equation for dipole loops implies that the maximum 
extent of the loop is given by 
\be
\xi_m q = 1 \, , 
\ee
where $q=q_m$ is the coordinate that labels the largest magnetic loop 
(which defines the boundary of the loop region). We drop the subscript 
from here on to simplify the notation. With $q$ specified, the critical 
values of the polar angle is determined by  
\be
q = 1 - \mu_m^2 \qquad {\rm or} \qquad \mu_m^2 = 1 - q \, . 
\label{qmudipole} 
\ee
Next we note that the value of $\xix$ is given by 
\be
\xix = (1 - \mu^2) / q \, . 
\ee
Using these results, we can write the integral of equation 
(\ref{volumeone}) in the form 
\begin{eqnarray}
V &=& {4\pi \over 3} q^{-3} \int_0^{\mu_m} d\mu \, 
\left[ \left( 1 - \mu^2 \right)^3 - q^3 \right] \nonumber \\ &= &
{4\pi \over 3} q^{-3} \left[ ( 1 - q^3 ) \mu_m - \mu_m^3 + 
{3 \over 5} \mu_m^5 - {1 \over 7} \mu_m^7 \right] \, . 
\end{eqnarray}
We can use equation (\ref{qmudipole}) to eliminate $\mu_m$ in 
favor of $q$, so that the expression becomes 
\begin{eqnarray}
V &=& {4\pi \over 3} q^{-3} (1 - q)^{1/2} \nonumber \\ &\times &
\left[ ( 1 - q^3 ) - (1 - q) + {3 \over 5} (1 - q)^2 
- {1 \over 7} (1 - q)^3 \right] \, , 
\end{eqnarray}
which simplifies to the form 
\be
V = {8\pi \over 105} q^{-3} (1 - q)^{3/2} 
\left[ 8 + 12 q + 15 q^2 \right] \, , 
\ee
Finally, we can write the volume in terms of the radial variable 
$\xi_m$ to obtain 
\be
V = {4 \pi \over 3} \xi_m^3 \cdot 
{2 \over 35} (1 - \xi_m^{-1} )^{3/2} 
\left[ 8 + 12 \xi_m^{-1} + 15 \xi_m^{-2} \right] \, . 
\ee
Note that the first factor is the total spherical volume enclosed
within the radius $\xi_m$, so that the second factor represents the
fraction of this fiducial volume that is enclosed by the loop region.

\section{Numerical Results} 
\label{sec:results}

This section presents the results of our numerical simulations, which
are divided into two classes. We first consider a set of simulations
that allow for outflow over the entire planetary surface. More
specifically, the goal of this initial set of models is to answer two
questions: [1] To what extent does the flow wrap around the planet to
the night side?, and [2] To what extent does the flow become
sufficiently powerful to open up the dipole field of the planet? In
order to isolate the effects of the magnetic field on the flow, these
models do not include rotation or tidal fields ({\bc see \citealt{trammell2014} for the impact of these effects on the flow structure}).  As shown below in
Section~\ref{sec:full_surface}, however, flow from the surface is
significantly suppressed from the night side of the planet because the
magnetic fields inhibit zonal flows.  As a result, we focus on the day
side of the planet and present a survey of parameter space that
considers only that hemisphere in Section~\ref{sec:hemisphere}.

\subsection{Preliminary Full-Surface Simulations} 
\label{sec:full_surface}

This subsection presents results from a preliminary set of simulations
that are designed to determine under what conditions the flow is able
to wrap around the planet and/or disrupt the planetary dipole. {\cb Note that for this setup, with the planetary dipole pointing in
the $\hat z$ direction, and the star located along the positive
$x$-axis, the problem is no longer axisymmetric. As a result, in order
to simulate the flow with a 2D grid, we must make additional
approximations. The flow is considered to be `pseudo-axisymmetric', where each cell is {\it locally} forced to have flow properties such that $\partial_\phi=0$ and there is no global requirement of axisymmetry, meaning we cannot include rotation (by performing a 2.5D simulation) which would introduce a non-physical shear along the poles. This obviously represents a restrictive situation and will only represent reality in the limit where the azimuthal flow is suppressed by the
magnetic field (as is the case here). Of course, once the outflow from
night-side of the planet is sufficiently small, the
numerical treatment reduces to that of the simpler, day-side-only 
simulations. Thus, these simulations allow us to place constraints on what magnetic field strengths one must have before the flow can truly be well approximated an axisymmetric day-side only simulation, which we present in Section 5.2. Such a condition is implicitly assumed in previous studies \citep[e.g.][]{adams2011,trammell2011,trammell2014} which we use these simulations to validate.   We  emphasise that in cases where there is a large scale azimuthal
flow (situations where field lines could start on the day-side and end
on the night-side, or vica-versa), then 2D simulations of this kind
cannot be used and only full 3D simulations are appropriate.}

\begin{figure*} 
%\figurenum{2}
\centering
\includegraphics[width=\textwidth]{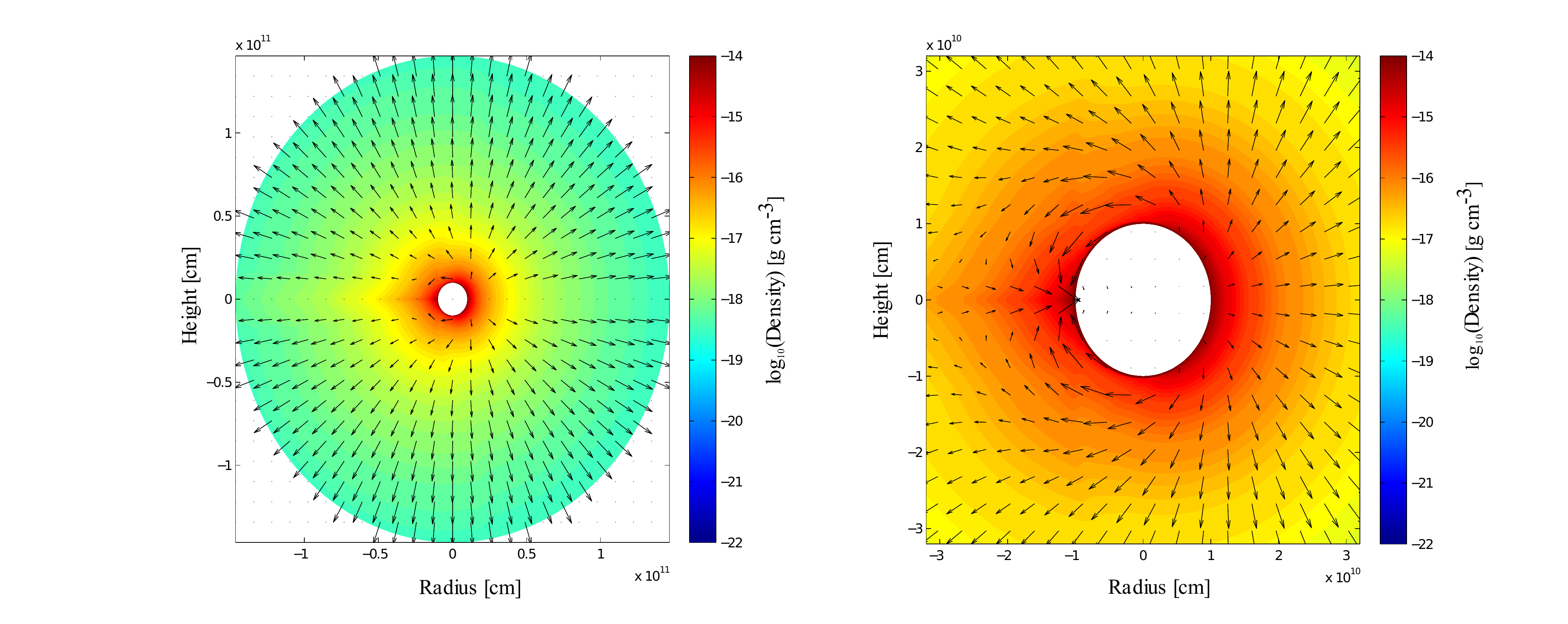}
\caption{Outflow solutions including both the day and night sides 
of the planet with no magnetic field. The colour map shows the density
and the vectors show the velocity field. The left-hand panel shows the
full simulation domain and the right-hand panel shows a zoom-in on the
planet. This model uses high levels of UV flux ($F_{UV}$ = $10^6$ erg 
cm$^{-2}$ s$^{-1}$). The star is located along the positive $x$-axis. 
Note that the outflow can be launched from all longitudes of the
planet, including the night side, in contrast to the case with a
magnetic field (compare with Figure \ref{fig:twopimagnet}).}
\label{fig:twopihydro} 
\end{figure*}   

\begin{figure*} 
%\figurenum{3}
\centering
\includegraphics[width=\textwidth]{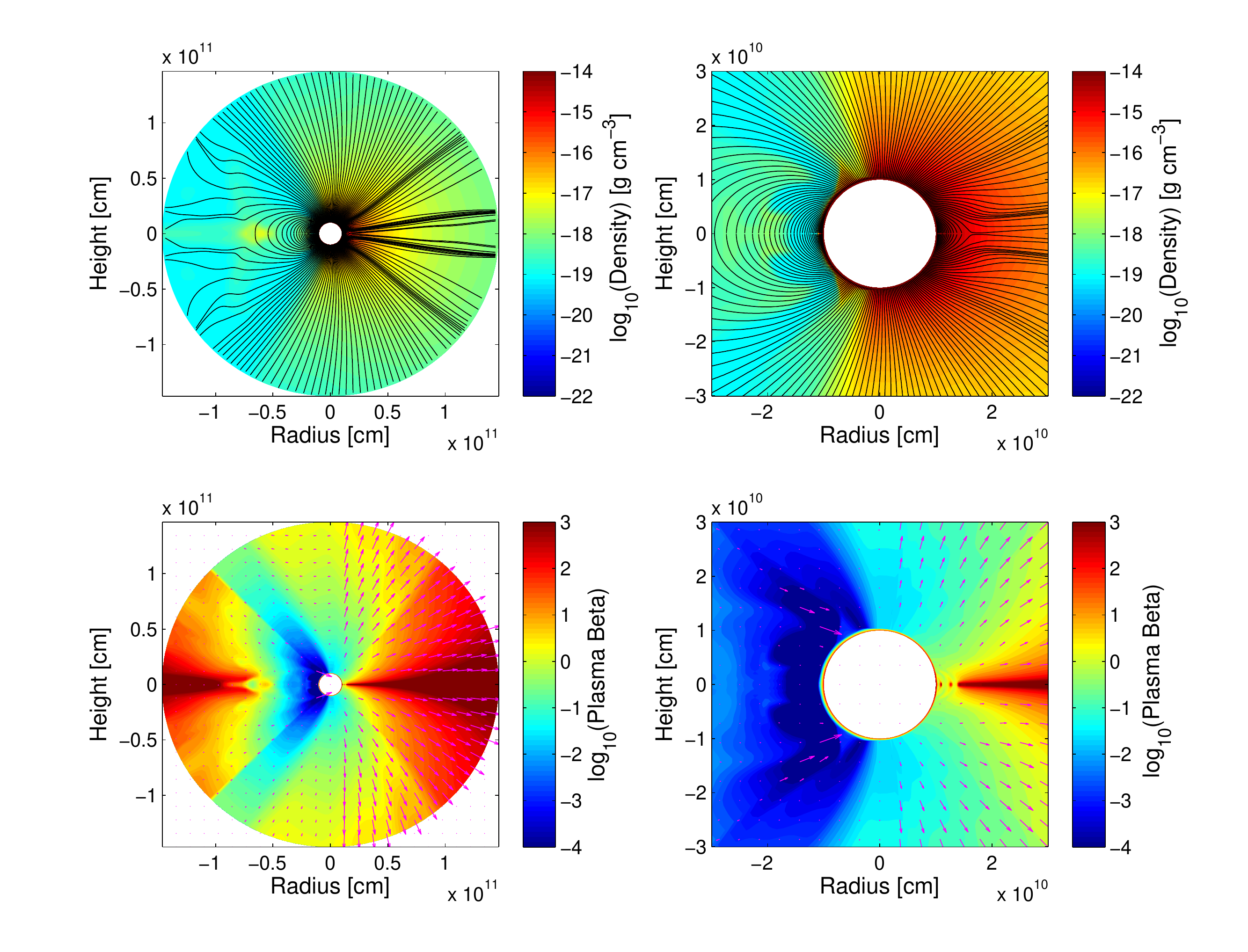}
\caption{Outflow solutions including both the day and night sides 
of the planet with a moderate magnetic field strength on the surface
($B_P$ = 0.3 gauss and $\bratio=0.0$) and UV flux of $F_{UV}=10^{6}$
erg~s$^{-1}$~cm$^{-2}$. The top panels show the density and magnetic
field structure; the bottom panels show the velocity structure and
Plasma beta. The left-hand panels show the full simulation domain and
the right-hand panels show a zoom-in on the planet. The star is located
along the positive $x$-axis. Note that the outflow is primarily
confined to the day side of the planet. }
\label{fig:twopimagnet} 
\end{figure*}  

These simulations use a Jupiter mass planet with radius $R_P$ =
$10^{10}$ cm. The magnetic field strength on the planetary surface is
taken to be $B_P$ = 0, 0.3, and 3 gauss.  The models are run with a
background stellar magnetic field that is aligned with the pole of the
planet and hence its dipole magnetic field. This configuration, which
is consistent with the analytic study of Paper I, is specified by the
parameter $\bratio$, which is defined by equation (\ref{betadef}). The
value of $\bratio$ sets the ratio of the background field to that on the
planetary surface. Here we take $\bratio$ = 0, 0.003, and 0.03.  The UV
flux from the star is chosen to have values from the high end of the
expected range, namely $F_{UV}$ = $10^5-10^6$ erg cm$^{-2}$ s$^{-1}$,
so that the flow becomes highly ionized and nearly isothermal with
temperature $T=10^4$ K.

As shown in Figure \ref{fig:twopihydro}, the flow wraps around the
planet in the absence of a planetary field, i.e., the outflow can
originate from essentially all longitudes. In addition, the outflow
becomes nearly radial at the substellar point.  In contrast, as shown
in Figure \ref{fig:twopimagnet}, the presence of even a moderate
magnetic field shuts down the outflow on the night side of the planet.
The simulation illustrated by Figure \ref{fig:twopimagnet} corresponds
to a relatively weak planetary magnetic field ($B_P=0.3$ gauss) and a
large UV flux ($10^6$ erg cm$^{-2}$ s$^{-1}$). Most of the expected
regime of parameter space corresponds to stronger planetary fields and
lower UV fluxes; we expect changes in both quantities to allow even
less heat transport to the night side of the planet. As a result, for
magnetically controlled flow, only the day side of the planet supports
outflowing streamlines. This complication reduces the expected
planetary mass outflow rates by a factor of $\sim2$.  For the main
survey of parameter space (see the following subsection), we thus
confine the simulations to the day side of the planet.

The outflow can only take place along open magnetic field lines.  
As discussed in previous sections, field lines can be open for two
reasons: [A] The pressure of the plasma at the planetary surface can
open up field lines, and [B] The background field of the star can open
up field lines (even in the absence of thermal pressure). In both
cases, the field lines are preferentially opened up along the poles.
We want to understand the extent to which these two effects are
operative. 

%\begin{figure} 
%%\figurenum{4}
%\centering
%\includegraphics[width=\columnwidth]{q_vs_kappa_2.eps}
%\caption{Latitude ($\theta_m$) of the last open magnetic field line, 
%and hence streamline, as a function of the parameter $\kappa=B_0^2/(8
%\pi P_0)$. The open circles show results for $\bratio=0$, the open
%squares for $\bratio=3\times10^{-3}$, and the stars are for
%$\bratio=3\times10^{-2}$. The dashed line shows the analytic result for
%$\bratio=0$ calculated in Section~\ref{sec:basicloop}.}
%\label{fig:qvskappa} 
%\end{figure}   

\subsection{Survey of Parameter Space}\label{sec:hemisphere}  

Given that the outflow is highly suppressed from the night side of the
planet, we henceforth limit our simulations to the day side {\cb by performing axisymmetric simulations}.  This
subsection presents results from a collection of simulations that
surveys the relevant parameter space.  Here we consider values of the
field strength ratio $\bratio$ = 0, $3\times 10^{-4}$, $1\times
10^{-3}$, $3\times 10^{-3}$, $1\times10^{-2}$ and $3\times10^{-2}$,
with planetary surface magnetic field strengths of $B_P$ = 0.5, 1.0,
4.0 and 10 gauss. The other important parameter is the UV flux, which
is taken here to have large values of $F_{UV}=10^{5}$ and $10^{6}$ erg
cm$^{-2}$ s$^{-1}$.

\begin{figure*}
\centering
\includegraphics[width=\textwidth]{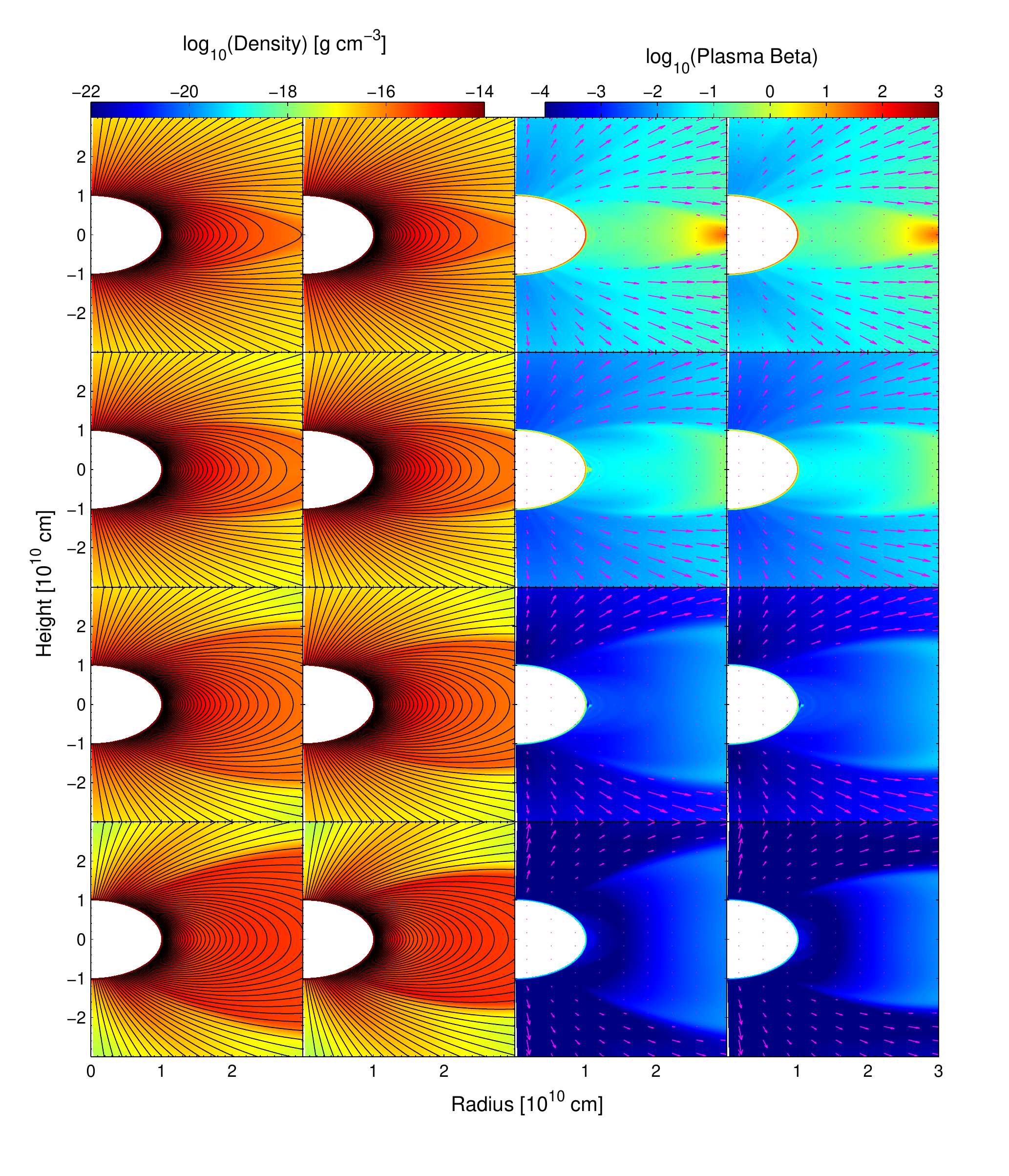}
\caption{Flow structure and field topologies for a subset of our 
simulated parameter space. The rows represent planetary magnetic
fields strengths of {\bc $B_P$ = 0.5, 1.0, 4.0 \& 10.0} from top to bottom. The first 
two columns show the density and magnetic field topology (first
$\bratio=0$, second $\bratio=3\times10^{-3}$); the last two columns show
the plasma beta and velocity structure (similarly: third $\bratio=0$,
$\bratio=3\times10^{-3}$). Note that these panels show a zoom-in on the
planet, whereas the full simulation domain extends out to $r$ = 
1.5$\times10^{11}$~cm (about 15 planetary radii). The star is located
along the positive x-axis.}
\label{fig:panel}
\end{figure*}

We demonstrate the effect of field opening from the pressure of the
flow and the background vertical field in Figure~\ref{fig:panel},
where we plot the flow topologies for simulations with a flux of
$10^{6}$ erg~s$^{-1}$~cm$^{-2}$, our four magnetic field strengths
(0.5, 1, 4.0 and 10 gauss from top to bottom) and $\bratio$ values of 0
and $3\times10^{-3}$ (left to right). The first two columns show
density and magnetic field topology while the second two columns show
the plasma beta and velocity structure. Comparing models with
different planetary field structures we see that at lower field
strengths (and hence higher plasma betas) the evaporative flow is able
to open out more and more closed field lines resulting in higher
mass-loss rates. As one increases $\bratio$ a similar result is seen
that the background field has opened out more field lines, resulting
in mass-loss from an increased surface area of the planet's
surface. {\bc We note for $B_P\lesssim 1$} gauss field opening due to the
flow dominates over the background field, but for $B_P\gtrsim 1$ the
number of opening field lines depends strongly on the strength (and
also topology) of the background stellar field. The fraction of the planetary surface that supports open
field lines can be defined by $\sin^2\theta_0$, where $\theta_0$ is
the polar angle of the last open field line (streamlines originating
at smaller angles are closer to the pole and hence open).

\begin{figure} 
%\figurenum{4}
\centering
\includegraphics[width=\columnwidth]{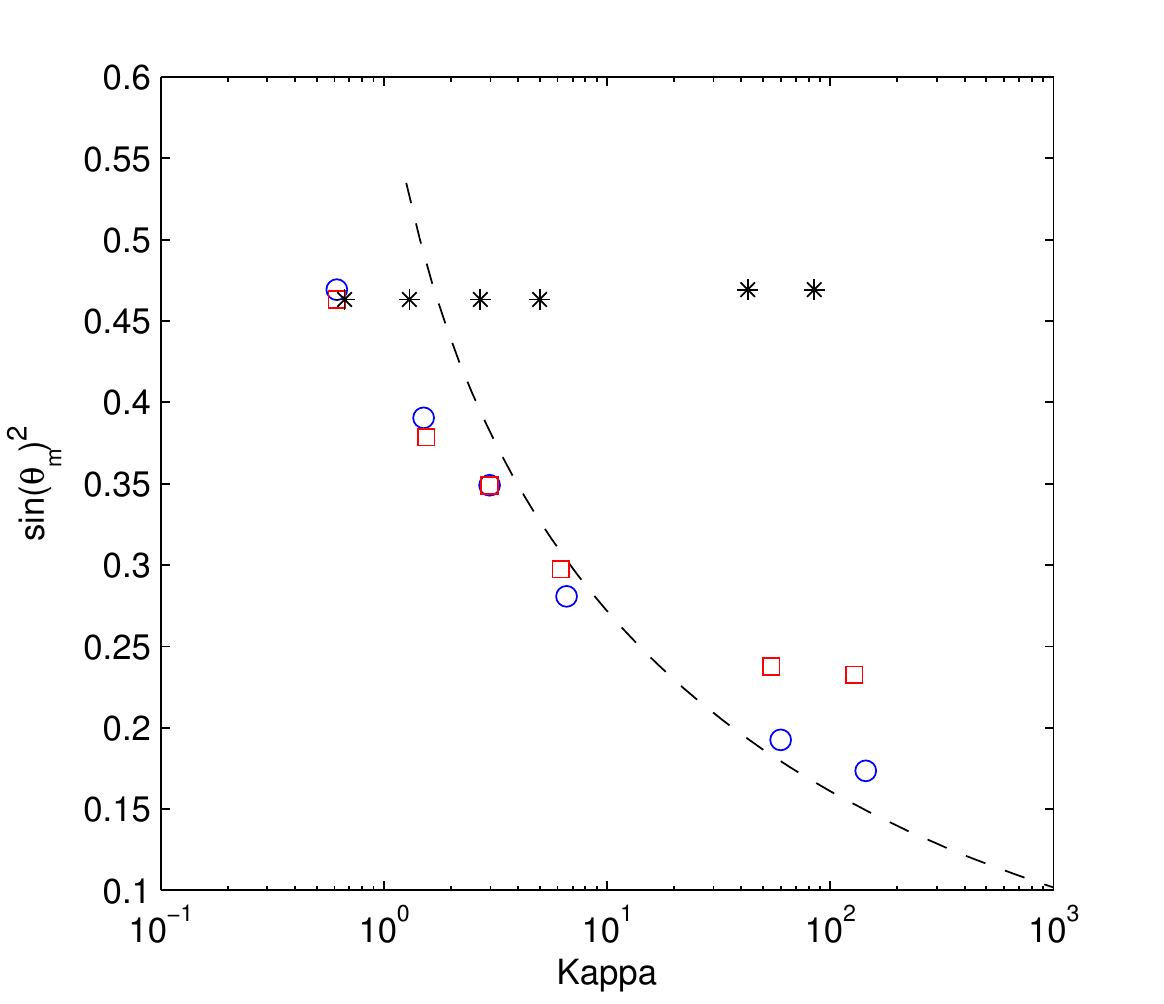}
\caption{Latitude ($\theta_m$) of the last open magnetic field line, 
and hence streamline, as a function of the parameter $\kappa=B_0^2/(8
\pi P_0)$. The open circles show results for $\bratio=0$, the open
squares for $\bratio=3\times10^{-3}$, and the stars are for
$\bratio=3\times10^{-2}$. The dashed line shows the analytic result for
$\bratio=0$ calculated in Section~\ref{sec:basicloop}.}
\label{fig:qvskappa} 
\end{figure}

Figure \ref{fig:qvskappa} shows the values of $\sin^2\theta_m=q_m$ as
a function of $\kappa$ for the simulations. Since the numerical
results for the pressure are not axisymmetric, and vary with latitude
as well, the value of $P_0$ used to determine $\kappa$ is taken to be
the latitudinally averaged pressure at the ionization front (which is
defined as the location where $X=0.9$).  The blue circles show the
results for a purely dipole field (no background stellar field, or,
equivalently, $\bratio=0$); the results closely follows the analytic
predictions, as shown by the dashed curve in the figure (where the
analytic result is shown for no rotation to be consistent with the
simulations). The red squares show the results for $\bratio$ = 0.003 and
the stars show results for larger $\bratio$ = 0.03. The fraction of the
planetary surface that supports open field lines depends on both the
pressure (defined via $\kappa$) and the background stellar field
(defined via $\bratio$).  However, for $\bratio = 0.01$, or larger, the
background stellar field provides the dominant contribution. Moreover,
even for these high UV fluxes, all plausible magnetic field strengths
will effectively control the structure of the planetary outflow.

\begin{figure} 
%\figurenum{5}
\centering
\includegraphics[width=\columnwidth]{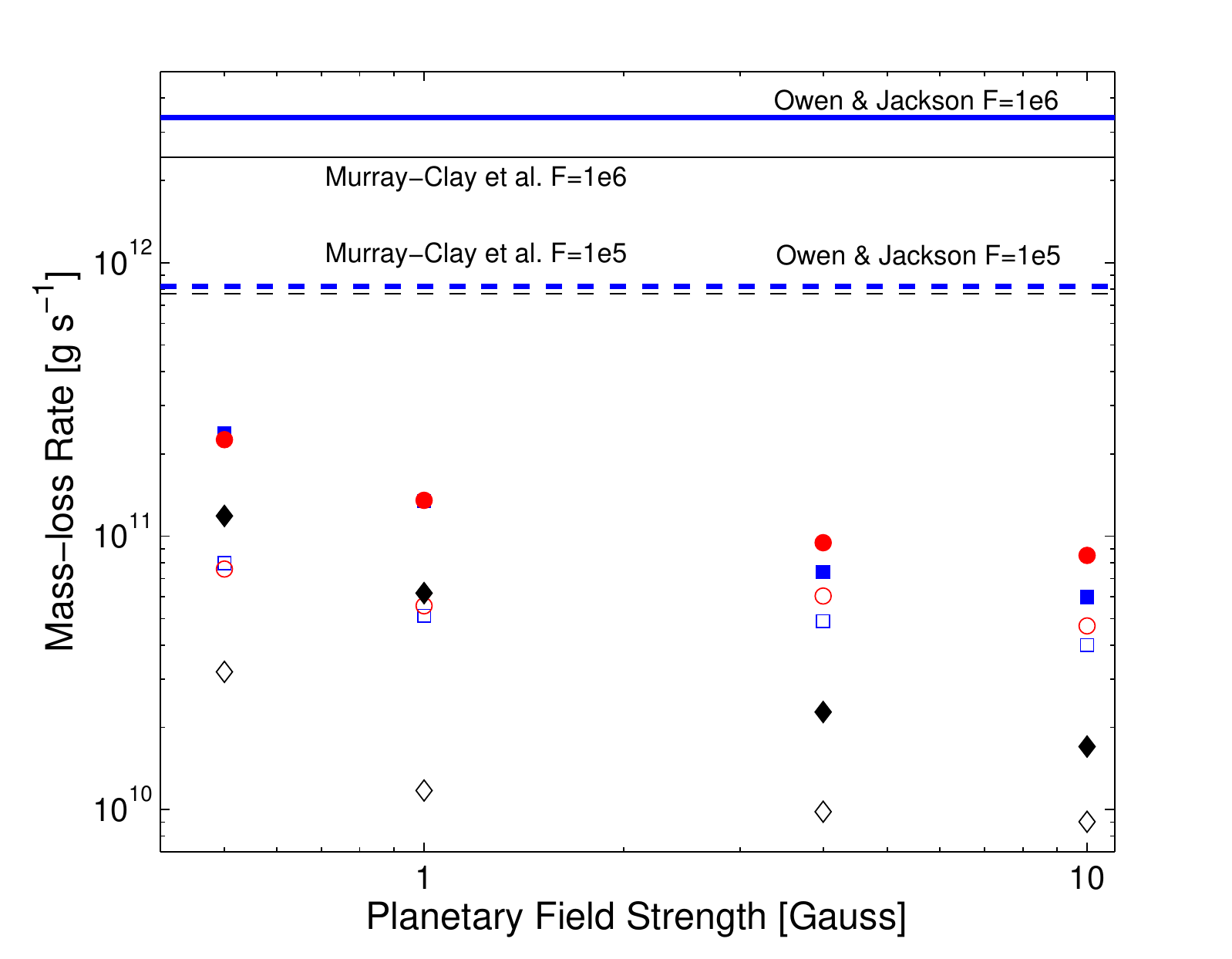}
\caption{Mass outflow rates as a function of magnetic field strength 
on the planet. The open (filled) symbols correspond to the lower
(higher) UV flux of $F_{UV}$ = $10^5$ ($10^6$) erg cm$^{-2}$ s$^{-1}$. 
The shapes of the symbols denote the value of the background stellar
field, defined via $\bratio$ = 0 (squares), 0.003 (circles), and 0.03
(diamonds). {\bc The horizontal lines denote the mass outflow rates for
planets with no magnetic fields (from: \citealt{mc2009}-thin/black and \citealt{oj12}-thick/blue), where we have
scaled these rates from 0.7~M$_J$ in \citet{mc2009} to 1~M$_J$ using
the scaling specified in Paper I. Note the \citet{oj12} rates also include a contribution from X-ray heating.}}
\label{fig:mdot} 
\end{figure}

Figure \ref{fig:mdot} shows the mass-loss rates from the simulations
plotted as a function of the magnetic field strength on the planetary
surface. Results are shown for a range of background stellar field
strength and hence a range of $\bratio$ = 0 -- 0.03, as well as the two
values of UV fluxes. For one set of the simulations (with $\bratio$ = 0.03),
the outflow does not reach a steady-state solution and we use a
time-averaged mass-loss rate (see below for further discussion).  For
comparison, the mass-loss rates are shown for purely hydrodynamic flow
\citep{mc2009,oj12}\footnote{Since \citet{mc2009} use a planet mass of
0.7~M$_J$ and we use a planet mass of 1~M$_J$, this comparison uses
the results of Paper~I to scale the results to our planet mass.}. The
net result is simple: The inclusion of the magnetic field results in a
clear suppression of the outflow rate, by approximately an order of
magnitude. This suppression is not unexpected, as magnetic planets
lose one factor of 2 because the night side flow is suppressed and
another factor of $\sim2-4$ because only a fraction of the field lines
are open. In addition to the overall suppression, the outflow rate
decreases with increasing magnetic field strength on the planetary
surface. On the right hand side of Figure \ref{fig:mdot}, the
combination of the stellar and planetary magnetic fields control the
geometry of the flow. On the left hand side of the figure, the fields
are weak enough that some (additional) field lines are opened up by
the plasma pressure, thereby increasing the outflow rate.

\begin{figure} 
%\figurenum{6}
\centering
\includegraphics[width=\columnwidth]{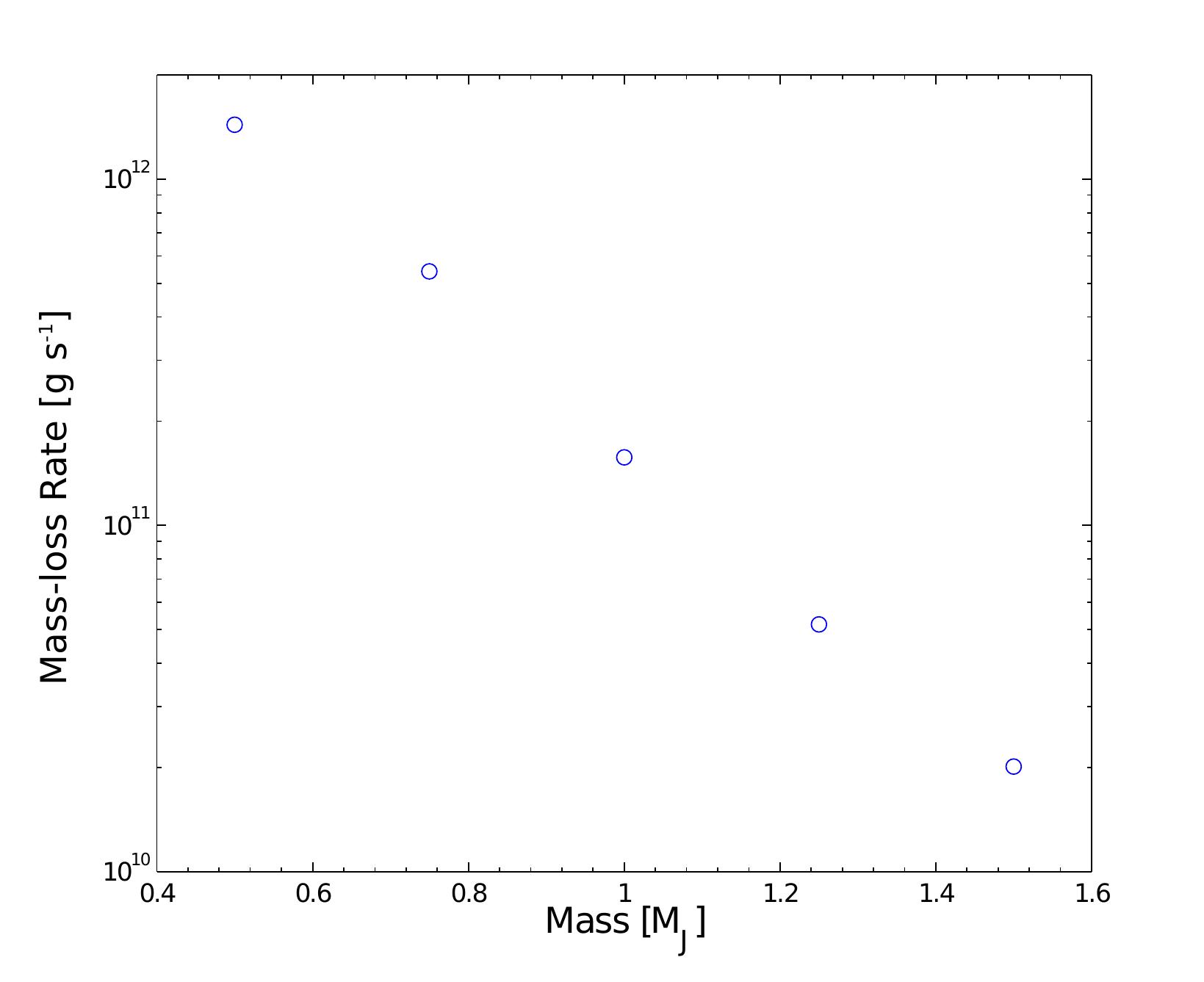}
\caption{Mass outflow rates as a function of planet mass. Results 
are shown for $R_P=10^{10}$ cm, $B_P=1$ gauss, $\bratio=0.003$, and 
$F_{UV}=10^6$ erg cm$^{-2}$ s$^{-1}$. The outflow rates show a nearly 
exponential decrease in $\dot M$ with increasing mass, in keeping 
with analytic expectations. }  
\label{fig:mdotmass} 
\end{figure}

\begin{figure} 
%\figurenum{7}
\centering
\includegraphics[width=\columnwidth]{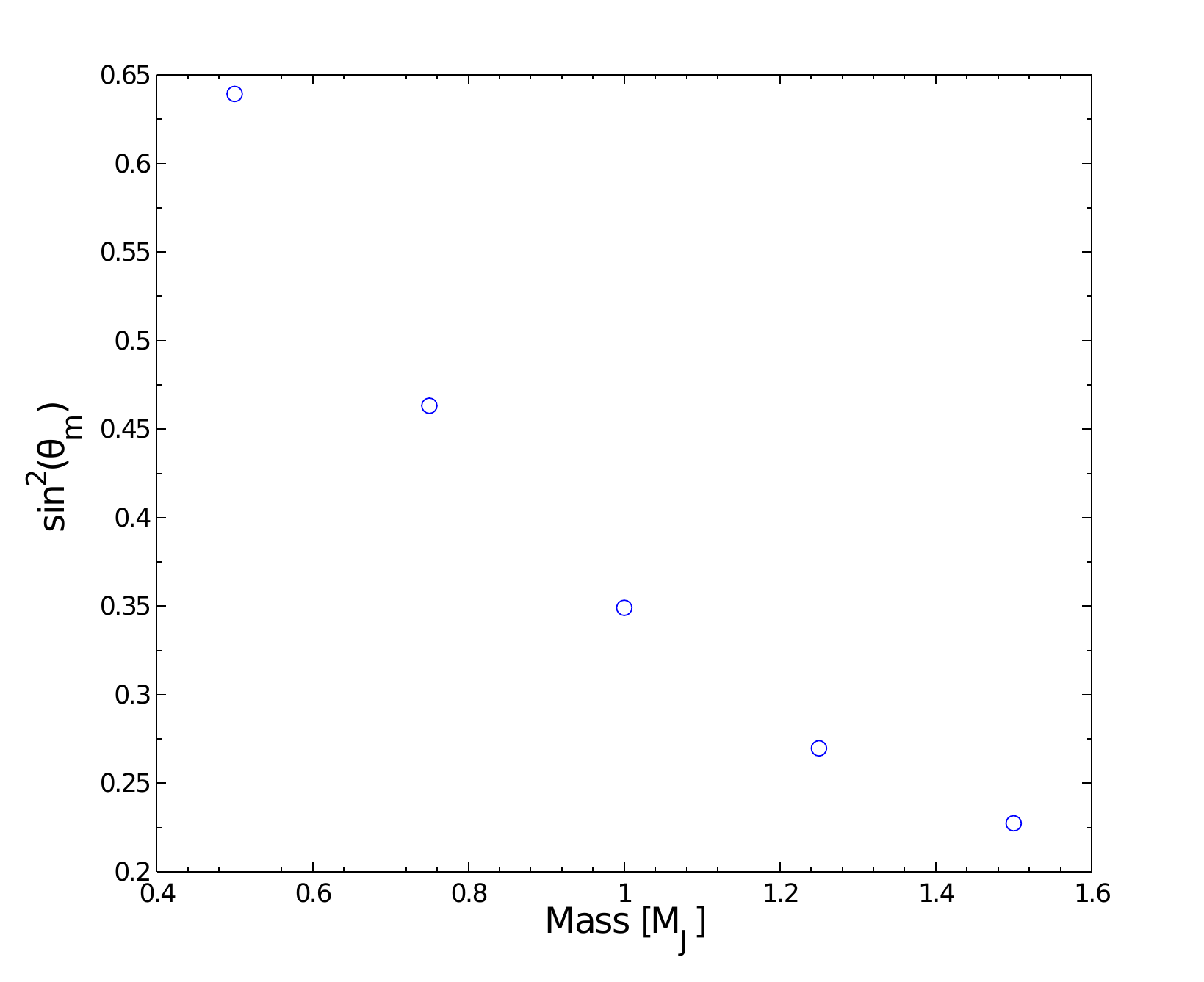}
\caption{Opening angle for outflow models as a function of planet mass. 
Results are shown for $R_P=10^{10}$ cm, $B_P=1$ gauss, $\bratio=0.003$,
and $F_{UV}=10^6$ erg cm$^{-2}$ s$^{-1}$ . }
\label{fig:qmass} 
\end{figure}

Figure \ref{fig:mdotmass} shows the mass-loss rates of the outflow as
a function of planet mass. For this set of simulations, the planetary
radius is held constant at $R_P = 10^{10}$ cm and the magnetic field
strength is fixed at $B_P=1$ gauss. The parameter $\bratio$, which sets
the strength of the background field due to the star, is also held
constant at $\bratio=0.003$.  Finally, the UV heating flux is fixed at a
constant value of $F_{UV} = 10^6$ erg cm$^{-2}$ s$^{-1}$.  These
simulations show that the mass loss rate ${\dot M}$ from the surface
decreases with increasing planet mass in a nearly exponential manner.
This general trend is consistent with the analytic prediction of Paper
I (see their equation [65]). In this case, however, the UV fluxes are
large, so that the low-mass planets have a larger fraction of their
surface accessible to outflow. This trend is illustrated in Figure
\ref{fig:qmass}, which shows the quantity $\sin^2 \theta_m$, where
$\theta_m$ is the polar angle of the last open field line (as discussed in section~\ref{sec:dipole}), as a
function of planet mass. At the planetary mass increases, the outflow
rate decreases, and fewer field lines remain open. Note that once the
outflow rate falls to a sufficiently low value, the fraction of open
field lines is determined primarily by the background field of the
star (through the parameter $\bratio$; see equation [\ref{fracfield}]).

\begin{figure}
\centering
\includegraphics[width=\columnwidth]{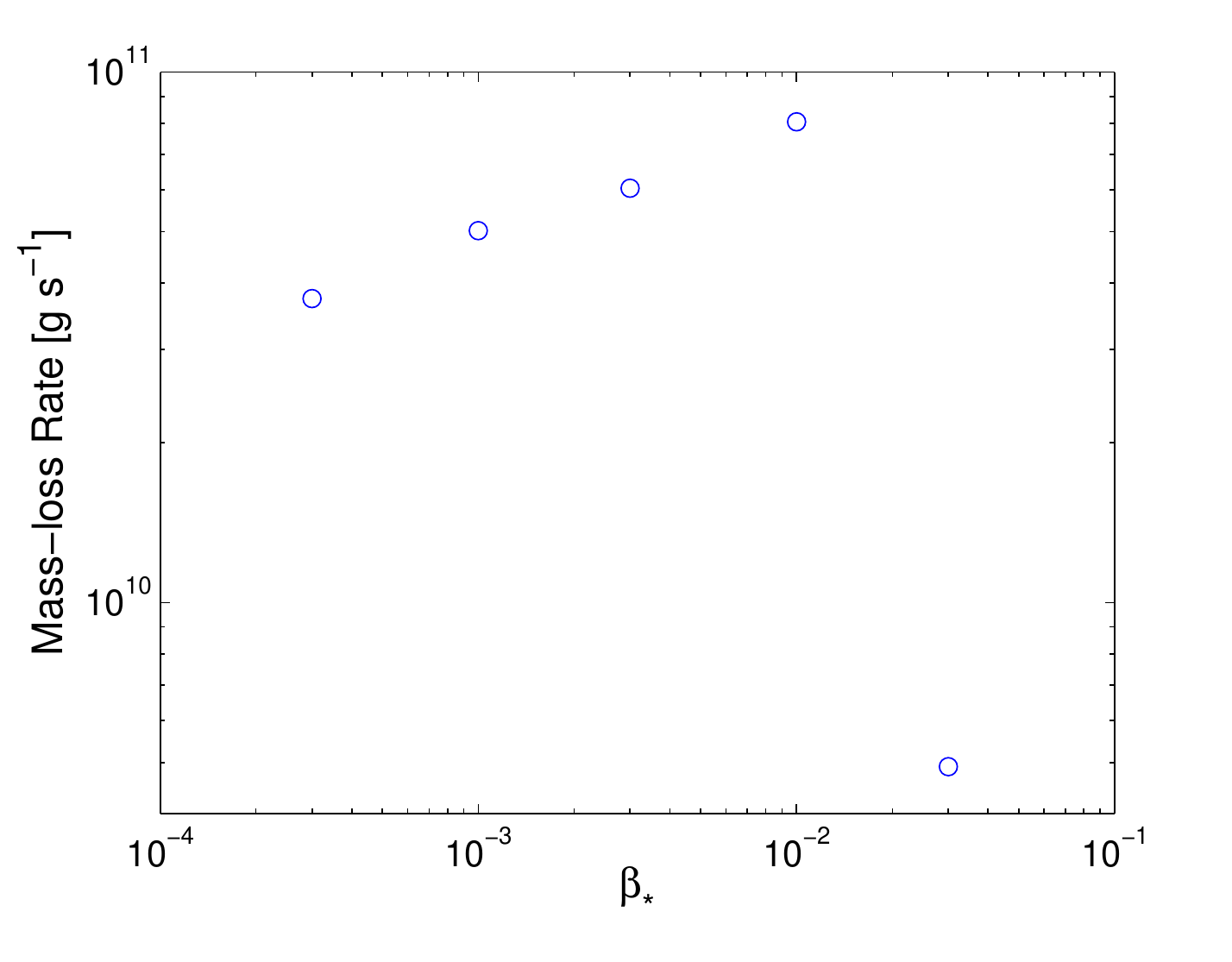}
\caption{Mass-loss rate as a function of the background field 
strength (encapsulated by the parameter $\bratio$). The set of 
simulations shown shown here uses $F_{UV}=10^{5}$ 
erg~s$^{-1}$~cm$^{-2}$, $B_P=4$ gauss, and $M_P=1$~M$_J$. 
The abrupt drop-off in the outflow rate occurs at large $\bratio$
because the flow cannot pass smoothly through the sonic point for 
all of the open field lines (see text). }
\label{fig:mdot_beta}
\end{figure}

Another interesting trend found in the simulations is that for
sufficiently large stellar contributions to the magnetic field (large
values of $\bratio$), the flow is suppressed further. This additional
suppression occurs even though the fraction of the planetary surface
that supports open field lines increases with $\bratio$. This trend is
shown in Figure~\ref{fig:mdot_beta}, where we plot the mass-loss rate
as a function of the parameter $\bratio$ for the set of simulations with
$F_{UV}=10^{5}$ erg~s$^{-1}$~cm$^{-2}$, $B_P=4$ gauss, and
$M_P=1$~M$_J$. Note that the mass loss rate rises slowly with
increasing $\bratio$, but then drops significantly for $\bratio>10^{-2}$.
Paper I predicts this type of behaviour: The initial rise occurs
because larger values of $\bratio$ lead to more open field lines; the
subsequent drop-off occurs because the flow cannot always pass
smoothly through the sonic point along all of the open field lines.
However, the suppression found here seems to be somewhat larger than
that indicated by Paper I and warrants further study.

\begin{figure}
\centering
\includegraphics[width=\columnwidth]{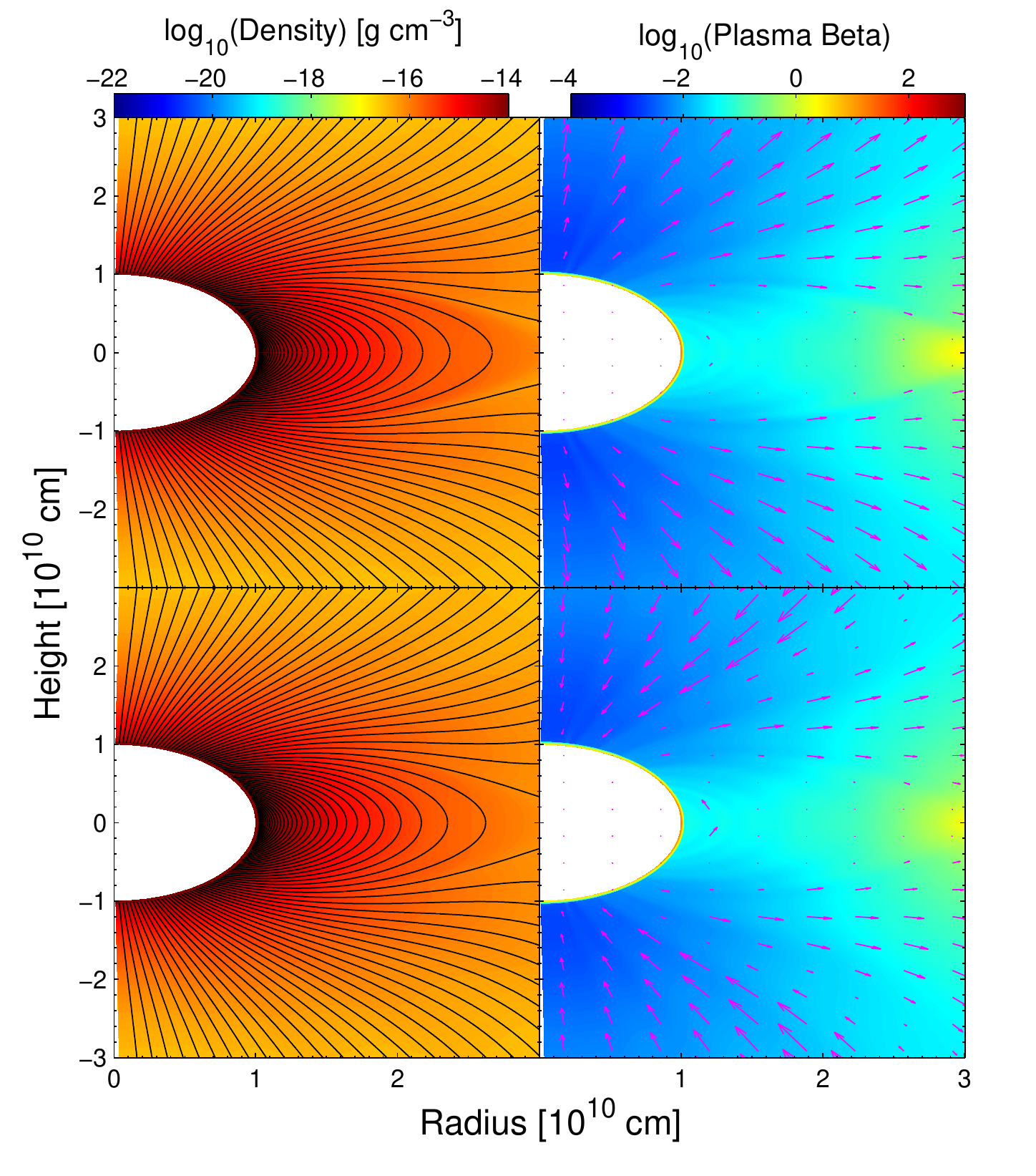}
\caption{Flow topology for the simulation with $B_P=1$ gauss,
$\bratio=3\times10^{-2}$ and $F_{UV}=10^{6}$ erg~s$^{-1}$~cm$^{-2}$. 
The top panels show outflow along all open field lines, while the
bottom panels show outflow along some open field lines and inflow
along others; the time between the two snapshots is $\sim6$~days. 
The left-hand column shows the density and magnetic field topology,
whereas the right-hand column shows the plasma beta and velocity
structure. The star is located along the positive x-axis. }
\label{fig:vari} 
\end{figure}

On a related note, for sufficiently large values of the parameter
$\bratio$ (which sets the strength of the background stellar field), the
simulations show that the flow does not always reach a steady state.
Instead, the mass loss rate varies with time. Along some field lines,
the flow is observed to alternate between the outward and inward
directions, i.e., it displays an apparently oscillatory behaviour.
This trend is demonstrated in Figure~\ref{fig:vari} for the simulation
with $B_P=1$ gauss, $\bratio=3\times10^{-2}$, and $F_{UV}=10^{6}$
erg~s$^{-1}$~cm$^{-2}$. For this case we find a variability time-scale
of $\sim 6$ days, and the Figure shows two snapshots of the flow
fields separated by 6 days. The magnetic field structure (depicted 
by the panels on the left side of the figure) does not change, 
consistent with magnetically controlled flow. On the other hand, 
flow along some field lines switches direction over the 6-day 
interval. 

Further, the flow generally does not reach the sonic point, but
instead remains subsonic.
%This behaviour thus corresponds to so-called breeze
%solutions, which we derive analytically in Appendix \ref{sec:breeze}.
The flow in this regime is much more difficult to simulate than the
case of steady, super-sonic winds. Because the flow is subsonic, and
even travels inward along some streamlines (field lines) at some
times, information can propagate from the outer boundary of the
simulation volume into the flow region. Therefore, the results -- in
particular the variability time-scale -- will be sensitive to the
conditions at the outer boundary, where the outer boundary conditions used are only exact for super-sonic outflow \citep{stone_hd}.

\section{Conclusion} 
\label{sec:conclude}

This paper has considered mass loss from Hot Jupiters in the regime
where the flow is magnetically controlled, including both numerical
simulations and supporting analytic calculations.  This section
presents a summary of our results (Section \ref{sec:sumresults}), a
discussion of their implications, and some recommendations for future
work (Section \ref{sec:discuss}).

\subsection{Summary of Results} 
\label{sec:sumresults} 

This work shows that essentially all outflows from Hot Jupiters are
expected to be magnetically controlled (provided that they support 
magnetic fields of moderate strength, $B_P \gtrsim 1$ gauss). This
conclusion follows both from analytic considerations (see Section
\ref{sec:parameters}) and from detailed numerical simulations (see
Section \ref{sec:results}).  Our simulations considered the most
extreme cases, those with the largest expected stellar UV fluxes and
moderate planetary fields strengths.  Even in this regime, however,
the magnetic field lines guide the flow and experience a negligible
back reaction.

The inclusion of magnetic fields leads to suppression of the total
mass outflow rate in three different ways. The first reduction arises
because the field suppresses zonal winds on the planet, so that heat
is not efficiently carried from the day side of the planet to the
night side (compare Figures \ref{fig:twopihydro} and
\ref{fig:twopimagnet}). The net effect is to essentially shut off the
outflow from the night side of the planet and thereby reduce the total
outflow rate by a factor of two.

The next type of suppression arises because not all magnetic field
lines can be opened up, so that only a fraction of the planetary
surface gives rise to outflow (see Figure \ref{fig:qvskappa}). For
the case of no background magnetic field from the star, the heated
plasma must produce a greater pressure than the magnetic pressure and
only the field lines close to the pole are open (see equation
[\ref{fractherm}]).  In the presence of a background stellar field, a
larger fraction of the planetary surface supports outflow, but the
active region is (again) confined to the poles (see Paper I and
equation [\ref{fracfield}]). This effect reduces the overall outflow
rate by another significant factor (2 -- 10), which depends on the
strength of the planetary field, the background stellar field, and the
plasma pressure.

The third source of outflow suppression arises because not all open
field lines allow the flow to make a smooth transition through the
sonic point. The degree to which this effect reduces the outflow rates
depends sensitively on the magnetic field geometry and other factors.
This effect is most pronounced for strong background fields, which
also act to open up more field lines (again see Paper I). As a result,
stronger background fields lead to competing effects of more open
field lines (implying more outflow) and difficulty in making the sonic
transition (implying less outflow).

With the inclusion of the magnetic fields, the overall mass outflow
rates are thus significantly smaller than indicated by previous work. 
Although the results depend on system parameters, over the regime
considered here, outflow rates for planets with magnetic fields are
about an order of magnitude smaller than those from planets with no
magnetic fields (see Figure \ref{fig:mdot}), and hence (about) an 
order of magnitude smaller than the simple estimates like that of 
equation (\ref{mdotestimate}). 

The outflow rates decrease sharply with increasing planet mass, 
as expected. This trend is nearly exponential, as shown in Figure 
\ref{fig:mdotmass}, and in agreement with analytic expectations 
(see equation [64] of Paper I). 

Finally, our simulations show that the outflows have time dependent
behavior in some portions of parameter space (Figure \ref{fig:vari}).
Our working hypothesis is that the flow cannot make a smooth
transition through the sonic point in this regime, so that the flow
solutions must vary with time (thereby resulting in non-steady flow).
The general finding of non-steady flow under these conditions is
consistent with the analysis of Paper I. Since the flow tends to be
subsonic in this regime the outer boundary condition can influence the
nature of the flow. This complication must be addressed in future work
(see also the discussion below).

\subsection{Discussion and Future Work} 
\label{sec:discuss} 

This work poses a number of interesting issues. First, we reiterate
that planetary outflows are expected to be magnetically controlled,
even for relatively weak fields ($B_P \sim 1$ gauss) and enormous UV
fluxes from the star ($F_{UV} = 10^6$ erg cm$^{-2}$ s$^{-1}$).
Moreover, the nature of the outflows (including mass loss rates and
flow patterns) must ultimately depend on the geometry of the magnetic
field, including the background contribution from the star.  As a
result, in order to understand planetary outflows, much more work must
focus on the magnetic field configurations (see Section
\ref{sec:parameters} for a discussion of the range of possible
configurations).

On a related note, the problem of planetary outflows naturally divides
into two regimes, the launch of the outflow and the subsequent
propagation of the flow after it passes through the sonic point.
Whereas this paper focuses on the launch of the outflow, the second
part of the problem remains largely unexplored. After the outflow
leaves the immediate vicinity of the planet, the flow structure
depends on the details of the stellar wind, the stellar magnetic
field, and the interactions of these fields with the planet (the
dimensionless parameters that define the regimes of interest are
outlined in Section \ref{sec:parameters}). If the stellar magnetic
fields are sufficiently strong near the planet, the planetary fields
will connect up with the background stellar field, which will control
the outflow away from the planet surface. On the other hand, if the
stellar wind overwhelms the stellar magnetic field before reaching the
planet, then a magnetospheric structure, similar to that seen for
Earth and Jupiter in our Solar System, will develop. In either case,
however, the characteristics of the stellar wind and/or stellar
magnetic field will determine the eventual structure of the outflow.

To date, most of the work carried out on planetary outflows has
focused on steady-state flow. Nonetheless, the results of our
numerical simulations indicate that the flow can be time-dependent.
The (relatively brief) discussion of this paper focuses on
time-dependence that arises from the difficulty that the flow faces in
passing smoothly through the sonic point (as anticipated in the
analytic treatment of Paper I).  However, a full treatment of this
issue remains to be carried out, and the results must ultimately
depend on the background stellar wind and stellar magnetic field
configurations. Further, these time-dependent outflows can be
subsonic, resulting in so-called breeze solutions. Unlike the case of
transonic flows, where the launch of the outflow is largely decoupled
from such outer properties, information can propagate inward through
subsonic flows (toward the planetary surfaces) from large distances,
so the background environment of the planet must play a role. In
practice, this property implies that the outer boundary conditions can
affect subsonic outflows originating from the planet surface.

In addition to the problem of passing through the sonic point,
time-dependent flow can arise from other sources. One interesting case
is that of planets executing eccentric orbits, where the distance from
the star varies appreciably over time. The UV heating rate will thus
vary over the orbit and the strength of the outflow will depend on
time.  For this configuration, one can calculate the thermal time
scale of the outflow, i.e., the time required for the UV heating flux
to provide the thermal energy of the outflow within the sonic surface, 
which is comparable to the kinetic energy. To leading order this time 
scale can be written in the form 
\be
t_{\rm th} = { \int \rho v^2 dV \over \pi R_P^2 F_{UV}} \approx 
{ {\dot M} \as \over \pi R_P F_{UV}} \approx 10^3 - 10^4 \, {\rm s}\,, 
\label{khtime} 
\ee
where we have used a moderate UV flux of 1000 erg cm$^{-2}$ s$^{-1}$
and we ignore factors of order unity. Although the time scale will
vary substantially from system to system, typical values range from 20
minutes to a few hours. The time required for the outflow to change
its properties can thus be much shorter than the orbital period,
thereby allowing for the possibility of observing time-dependent
outflows in eccentric systems. Note that the thermal time scale of
equation (\ref{khtime}) is roughly comparable to the sound-crossing
time of the subsonic region $t_s = few \times R_P/\as$. This convergence
arises because the mechanical luminosity of the outflow is roughly 
the same as the rate of energy absorption from UV radiation. In 
practice, however, some losses occur and $t_s > t_{\rm th}$.

Thus far, planetary outflows are only observed in two systems,
although additional observations should be forthcoming. In addition to
measuring the outflow rate, however, additional observational
signatures must be developed. In order to solve for the flow
properties, one must take into account the detailed heating and
cooling mechanisms.  This information, in turn, determines the
possible emission lines produced in the outflow. If one solves for the
chemistry of the outflow region, between the planetary surface and the
surface where the flow becomes optically thin at UV wavelengths, then
the absorption features can be determined. Finally, we (again) note
that the magnetic field controls the flow over the entire region where
the outflow can be observationally detected. As a result, the UV
should display a polarization signal (although such measurements are
difficult; see, e.g., \citealt{sloane} for further discussion).

Mass loss from the planet can ultimately affect the spin rate of the
planet (in the absence of other torques). Since the mass loss is
asymmetric, taking place only on the day side of the planet, the flow
itself can carry away angular momentum. In addition, however, magnetic
torques associated with the magnetic stresses guiding the planetary
outflow will play a role \citep{weberdavis}. In this case, the
magnetic torques will generally be dominant, by approximately the
ratio of the magnetic pressure to the ram pressure of the outflow (see
Section \ref{sec:parameters}), i.e., by a large factor. Nonetheless,
both types of torques should be studied in the future, as they can
influence the planetary spin rate and perhaps even the orbital angular
momentum.

Finally, we note that the present study has focused on planets with
Jovian masses, where the escape speed from the surface is large enough
so that the outflow rates are relatively small. To put this statement
in context: Using the results shown in Figure \ref{fig:mdotmass}, a
1.0 $M_J$ planet has an outflow rate ${\dot M}\approx10^{11}$ g/s,
which turns out to be about 0.0016 $M_J$/Gyr. With these low mass loss
rates the planetary mass will not change much over its lifetime. For
planets with smaller masses, however, the outflow rates can be large
enough to affect planetary masses. If we extrapolate the trend shown
in Figure \ref{fig:mdotmass} down to the mass $M_N$ of Neptune, the
mass loss rate is about 3.15 $M_N$/Gyr, large enough to make an
enormous difference. As the planet mass decreases, the outflow rates
increase and the amount of mass loss required to affect the planet
decreases.  As a result, we expect a type of cross-over mass such that
larger planets are only moderately affected by mass-loss and smaller
planets are efficiently evaporated down to their rocky cores or ocean
surfaces (a similar threshold has been suggested by \citealt{owenwu}).
This work indicates that the mass threshold will depend on the
magnetic field structure of these intermediate-mass planets. A crucial
question is thus whether or not close-in planets with mass comparable
to Neptune will support moderately strong magnetic fields.

\medskip 
\section*{Acknowledgments:}We are grateful to the referee for a constructive report that helped improve the manuscript.
We would like to thank Marcelo Alvarez, Barbara Ercolano and Garrelt
Mellema for helpful discussions. The numerical calculations were
performed on the Sunnyvale cluster at CITA, which is funded by the
Canada Foundation for Innovation. We are grateful for the hospitality
of both CITA and the University of Michigan for visits that helped
facilitate this collaboration.

%\newpage 

\appendix 

\section{Radiative Transfer and Ray Tracing} 
\label{sec:radtransfer} 

Our problem possess the two challenging aspects commonly encountered
in radiation-hydrodynamics. Firstly, our ray-tracing scheme must be
performed in a causal manner (\citealt{mellema06} as the attenuation
is strongly linked to the ionization structure, which in turn depends
on the level of attenuation). Secondly the symmetry of the radiation
field (plane-parallel) does not match the symmetry of the planet's
atmosphere ($\sim$ spherical). One could solve the second problem by
performing the calculations on a Cartesian grid; however, setting up a
spherical planetary atmosphere in Cartesian grid is problematic: it
requires a large number of cells in the vicinity of the planet; 
{\sc zeus}'s directionally-split MHD algorithm may result in spurious
numerical artefacts resulting from strong gradients inevitably
mis-aligned with the grid, and finally it makes defining a boundary
condition below the planet's atmosphere difficult.
 
Therefore, for maximum accuracy at minimal computational cost in the
MHD scheme we choose to evolve our problem on a spherical grid and
perform plane parallel ray-tracing through a spherical grid. Casting a
single ray for every cell is prohibitively expense so we develop a
hybrid-characteristics scheme. Such a scheme (for the reverse problem
of spherical ray-tracing on a Cartesian grid) has been shown to be
accurate, efficient and parallelisable and scalable over {\sc mpi } by
\citet{hybrid_char}. The essence of the scheme is to decompose the
grid into small `blocks'. Within each of these blocks one performs a
long characteristics ray-tracing calculation for every cell in the
block. In between the blocks the optical depths are interpolated onto
the new rays using the method of short-characteristics; as such it is
not as diffusive as a fully short-characteristics based scheme while
still being computational feasible.

\begin{figure*}
\centering
\includegraphics[width=\textwidth]{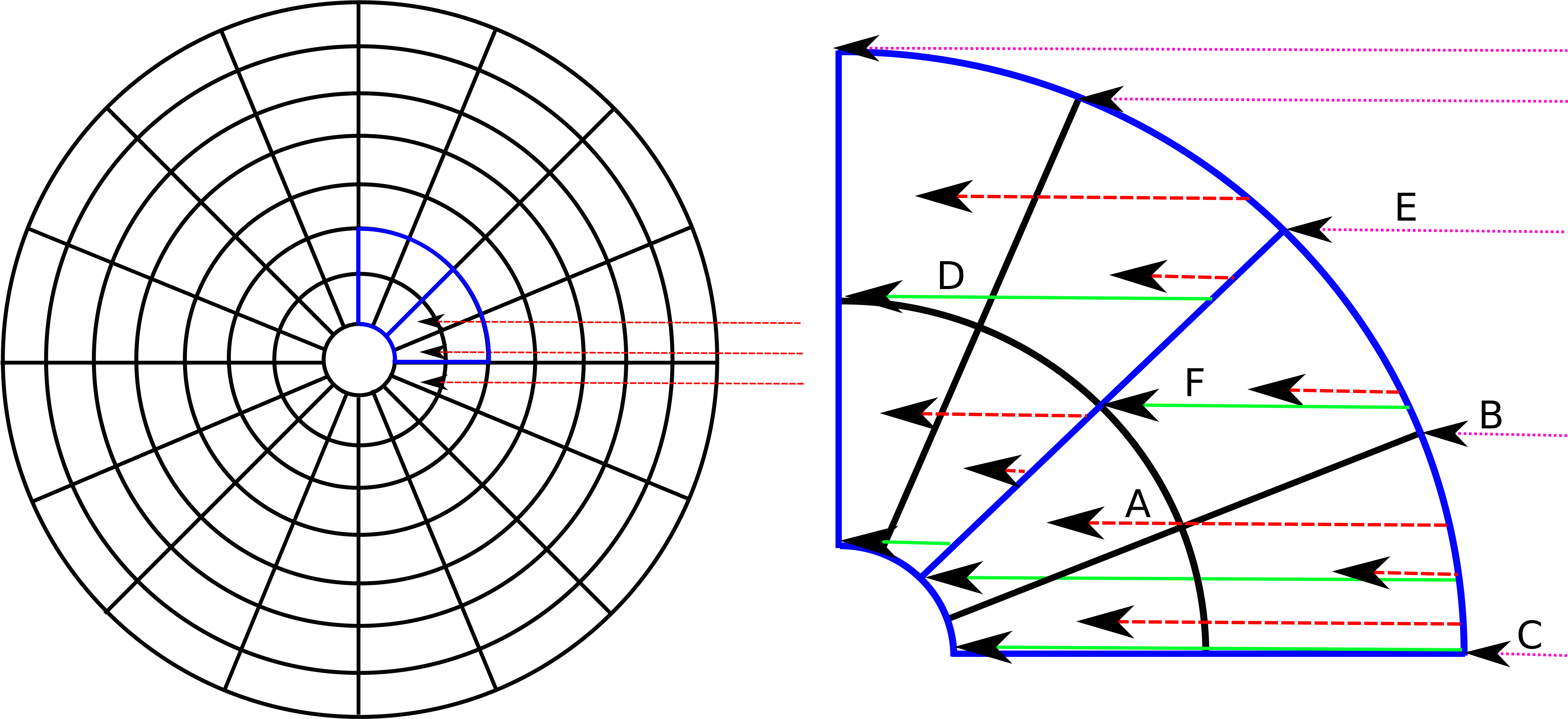}
\caption{Schematic diagram showing how the hybrid-ray tracing scheme  
works in practice for plane-parallel radiative transfer on a spherical
grid. The left hand panel shows how we decompose the grid into smaller
`blocks', indicated by the blue highlighted grid structure. The
right-hand panel shows a zoom-in of an individual block and shows all
of the associated rays present in the calculation. The dotted rays are
those originating in previous blocks, which are required for the
interpolation scheme used to calculate the starting optical depths for
the rays in this block. The dashed rays show the rays calculated for
every cell in this block. The solid rays indicate all of the extra rays
needed for the interpolation in the next block to calculate the
starting optical depths. The letter labels are used in the text to
describe the interpolation scheme.}
\label{fig:ray_tracing}
\end{figure*}

We follow \citet{hybrid_char}, such that within each block we perform
a ray-tracing calculation for each cell and each cell corner on which
short-characteristic interpolation takes place. The ray-structure is
schematically shown in Figure~\ref{fig:ray_tracing}. Thus, referring
to the right-hand panel of Figure~\ref{fig:ray_tracing}, the optical
depth at the beginning of ray `A' would be determined by bi-linear
interpolation on the radial cell boundary using the optical-depths at
the ends of rays `B' and `C'; furthermore, the optical depth at the
beginning of ray `D' would be determined by bi-linear interpolation on
the angular cell boundary using the optical depths at the end of rays
`E' and `B'.  Such a scheme is particularity amenable to
parallelisation in MHD codes that are already parallelised in a
block-domain-decomposition manner (such as {\sc zeus}). If the
decomposed blocks for the ray-tracing scheme fit evenly within the
blocks of the MHD scheme then non-local transfer of an information
along an individual ray is not required between CPU's. The ray-tracing
is then performed in a casual manner where we step through the blocks
(and cells within a block) such that the ionization structure of all
previous cells any given ray intercepts is calculated before the
ionization structure of any given cell is calculated
\citep[e.g.][]{mellema06}.

In all our calculations we decompose the 2D grid into `blocks' of size
$8\times8$, which we find gives a good balance between accuracy and
performance. It is unclear whether such a scheme is suitable for
implementation in a 3D code, or whether a ray-splitting approach is
more appropriate \citep[e.g.][]{moray} and will be investigated in
future work.

\section{Numerical Approach}\label{sec:numerical}

In order to study the evaporation of hot Jupiter atmospheres we have
developed an ionization radiative transfer method for the well known
{\sc zeus-MPv2} astrophysical MHD code
\citep{stone_hd,stone_mhd,hayes06}. The {\sc zeus} code is a robust
and well tested MHD code, where the magnetic field is evolved using
`constrained transport' \citep{ct} which preserves ${\bf
  \nabla}\cdot{\bf B}={\bf 0}$ to machine precision provided the
magnetic field is initialised with ${\bf \nabla}\cdot{\bf B}={\bf
  0}$. We choose to reconstruct fluxes at cell boundaries in a second
order fashion using a Van-Leer limiter and the artificial viscosity is
chosen such that discontinuities in the flow are smoothed over
approximately two cells ($q=2.0$). {\cb In several of the update sub-steps it was necessary to replace the finite-difference operators with appropriate finite-volume operators. In particular, this was necessary for the simulations with both the day and night-side \citep[see][Apendix B2]{hayes06}.} In order to model the evaporation
of hot Jupiters due to Ionizing EUV radiation we need to include a
radiative transfer scheme that solves for the radiation field in a
given cell, the ionization structure and gas temperature. Our scheme
is separated in two main components: a scheme that solves for the
ionization and thermal structure of a given cell and a ray-tracing
scheme that captures the transport and attenuation of EUV photons
(previously discussed in Appendix~\ref{sec:radtransfer}).

The evolution of the ionization structure is governed by
equation~(\ref{eqn:Ion_balance}). We follow {\sc zeus}'s natural
structure and solve equation~(\ref{eqn:Ion_balance}) using operator
splitting, by splitting it into a `source' step and a `transport'
step. In the source step we ignore the advection term and simply
solve the equation 
\begin{equation}
\frac{{\partial}X}{{\partial}t}=
(1-X)(\Gamma +n_eC)-Xn_e\alpha_r\,. 
\label{eqn:ion_source}
\end{equation}
In the transport step we account for the passive advection of
electrons, ions and neutral species. {\sc zeus} provides a built-in
feature to perform this update and this is performed as
\citep[see][for details]{hayes06}, 
\begin{equation}
\frac{d}{dt}\int_V\rho_idV=-\oint_{\partial V}
\rho_i{\bf v}\cdot d{\bf S}\,,
\label{eqn:ion_adv}
\end{equation}
where $\rho_i$ is the mass density of the advected species, $V$ is the
cell volume, $d{\bf S}$ is the cell surface area element and ${\bf v}$
is the gas velocity. Operationally, we update
equation~(\ref{eqn:ion_source}) after the usual {\sc zeus} body-force
and artificial viscosity steps as well as updating the gas temperature
and update equation~(\ref{eqn:ion_adv}) after the {\sc zeus} constrained
transport and transport steps.

\subsection{Ionization and Thermal balance}

We make the common On-The-Spot approximation, i.e., we assume that the
recombinations to the ground state are locally reabsorbed
\citep[e.g.,][]{spitzer,mellema06,ivine}.  As a result, for an
ionizing flux entering a cell we can update the ionization/thermal
structure of the cell, along with calculating the ionizing flux
leaving the cell (required to give the ionizing flux entering the next
cell). Our ionization scheme is based on the {\sc c2Ray/doric} scheme
\citep{mellema06} and our thermal update is similar to that used in
the {\sc iVine} code \citep{ivine}. The {\sc c2ray} scheme solved many
of the problems arising from ionization radiative transfer in
hydrodynamic simulations \citep{mellema06}, particularly problems
associated with photon conservation, sharp ionization fronts, and the
propagation of rapidly moving R-type fronts (not an issue for our
problem). Because the {\sc c2Ray} scheme is described in detail in
\citet{mellema06}, only the basics are presented here. For simplicity,
we only consider hydrogen so that the evolution of the ionized
fraction ($X$) is given by  
\begin{equation}
\frac{\partial X}{\partial t}=(1-X)(\Gamma +nXC)-nX^2\alpha_r\,. 
\label{eqn:ionization}
\end{equation}
In the On-The-Spot approximation the recombination rate is simply the
Case B recombination coefficient ($\alpha_b$) given by {\bc \citep[e.g.][]{mellema06}}
\begin{equation}
\alpha_b=2.59\times10^{-13}\mbox{~cm$^3$~s$^{-1}$~}
\left(\frac{T}{10^{4}\mbox{~K}}\right)^{-0.7} \,. 
\end{equation}  

Equation~(\ref{eqn:ionization}) can be solved iteratively using the
method suggested by \citet{SchmidtVoigt87}, where one takes $\Gamma$,
$n_e=Xn$, $C$, and $\alpha_b$ to be constant over a time step such that 
equation~(\ref{eqn:ionization}) has the solution 
\begin{equation}
X(t+\Delta t)=X_{\rm eq}+\left[X(t)-X_{\rm eq}\right] 
\exp\left(-\frac{\Delta t}{t_{\rm ion}}\right)\,,\label{eqn:iterate}
\end{equation}
where we have defined 
\begin{equation}
t_{\rm ion}=\frac{1}{\Gamma +n_eC+n_e\alpha_b}
\end{equation}
and
\begin{equation}
X_{\rm eq}=\frac{\Gamma +n_eC}{\Gamma +n_eC+n_e\alpha_b}\,.
\end{equation}
The values of $\Gamma$, $n_en$, $C$, and $\alpha_b$ can then be
recalculated using the new ionization state and
equation~(\ref{eqn:iterate}) can be evaluated again and so-forth.  The
essence of the iteration scheme is to repeatedly solve
equation~(\ref{eqn:iterate}) where $t_{\rm ion}$ and $X_{\rm eq}$ are
replaced with {\it time-averaged} quantities, which are updated using
the previous ionization structure of the cell and the value at the
current iteration, until convergence is achieved and $t_{\rm ion}$ and
$X_{\rm eq}$ represent the correct time-averaged quantities. The
advantage of such an iteration procedure is it readily provides an
analytic formula for the time-average ionization fraction by
time-averaging equation~(\ref{eqn:iterate}).

During each iteration of equation~(\ref{eqn:iterate}), we need to know
the gas temperature. In reality this would require solving the thermal
balance equations iteratively as well. However, since in this study we
our restricting ourselves to the highest UV fluxes where the flow is
close to recombination equilibrium \citep{mc2009}, we can dramatically
simplify the thermal balance problem. Since fully ionized gas has a
temperature ($\sim 10^{4}$ K) we follow the method of \citet{ivine}
and adopt a gas temperature profile of 
\begin{equation}
T=XT_{\rm hot}+(1-X)T_{\rm cold}\,,
\end{equation}     
where we set $T_{\rm hot}=10^{4}$~K and $T_{\rm cold}=10^{3}$ K, where
our choice of $T_{\rm cold}$ is approximately the temperature of the
underlying bolometrically heated atmosphere (we note that this choice
makes little difference to our results provided that the scale height 
of the underlying atmosphere is much smaller than the planetary radius
\citealt{mc2009,oj12}). The internal energy of the gas is then found
via and ideal equation of state, such that
\begin{equation}
u=\frac{1}{\gamma-1}\frac{k_b(1+X)}{m_h}\rho T\,,
\end{equation}
where we choose $\gamma=5/3$. This parametrization is valid only if
the ionization front is small compared to the flow-scale; this
constraint restricts our investigation to the highest UV fluxes
experienced by hot Jupiters.

\subsection{Numerical Tests}

\begin{figure}
\centering
\includegraphics[width=\columnwidth]{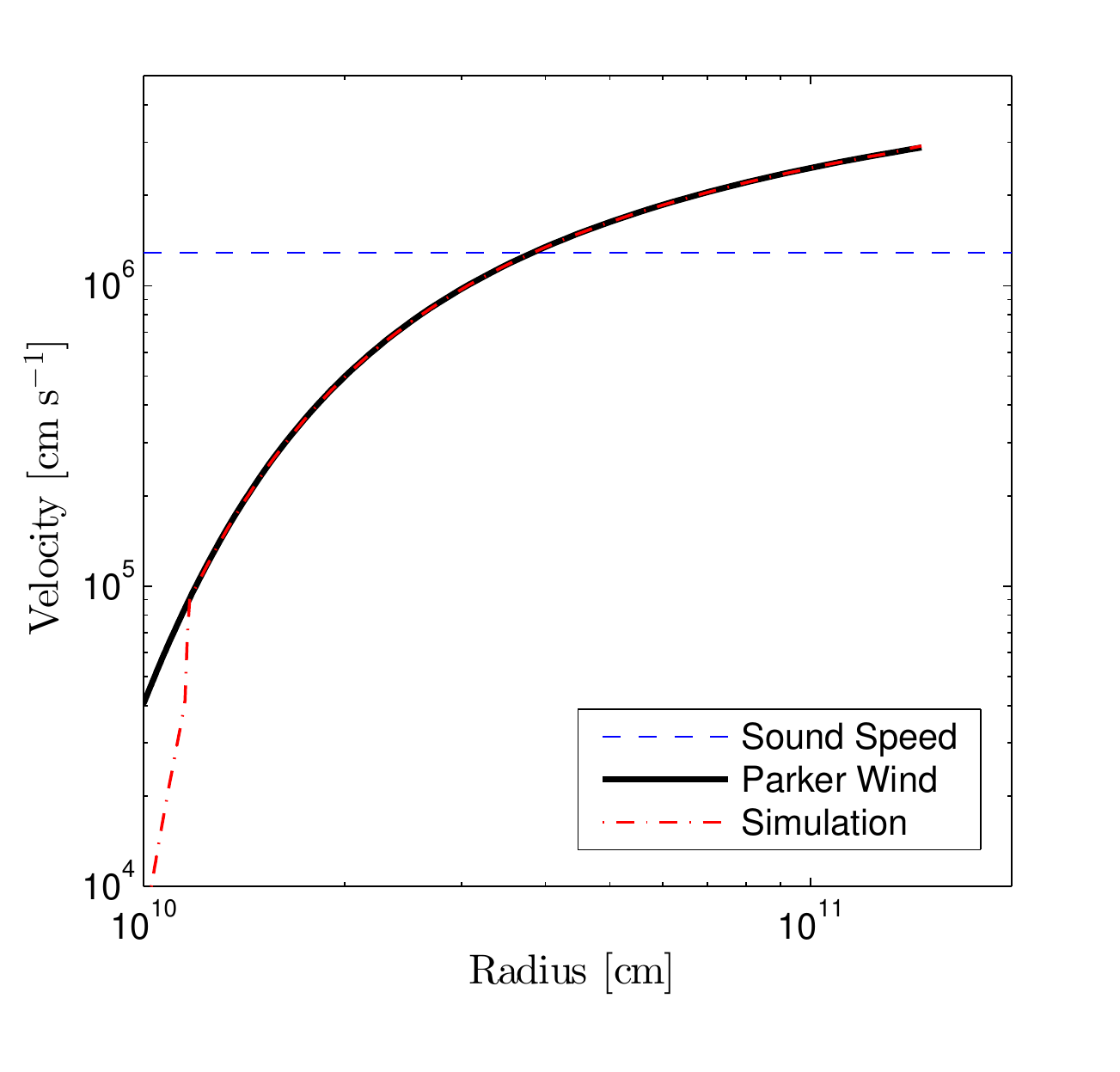}
\caption{Figure showing the results of the Parker wind test, i.e., 
the flow speed as a function of radius for both the numerical and
analytic solutions.  After the flow passes through the ionization
front, the velocity closely follows the analytic Parker wind solution 
(as expected). }
\label{fig:parker_test}
\end{figure}

Since we have not developed a new scheme from scratch, we restrict
this discussion to a small number of test problems aimed specifically
at our problem in hand. Thus, we perform 1D spherically symmetric flow
calculations without magnetic fields. In this case the problem can be
well approximated by the isothermal `Parker-wind' problem which posses
and analytic solution for the velocity structure of the flow
\citep{parkerspiral}. In Figure~\ref{fig:parker_test} we show the
velocity structure resulting from our code compared to the
`Parker-wind' solution for a 1~M$_J$ planet with a radius of
$r_p=10^{10}$~cm and UV flux of $10^{5}$ erg~s$^{-1}$~cm$^{-2}$. We
note that the flow on the upstream side of the ionization front should
match onto the Parker wind solution. 

\begin{figure}
\centering
\includegraphics[width=\columnwidth]{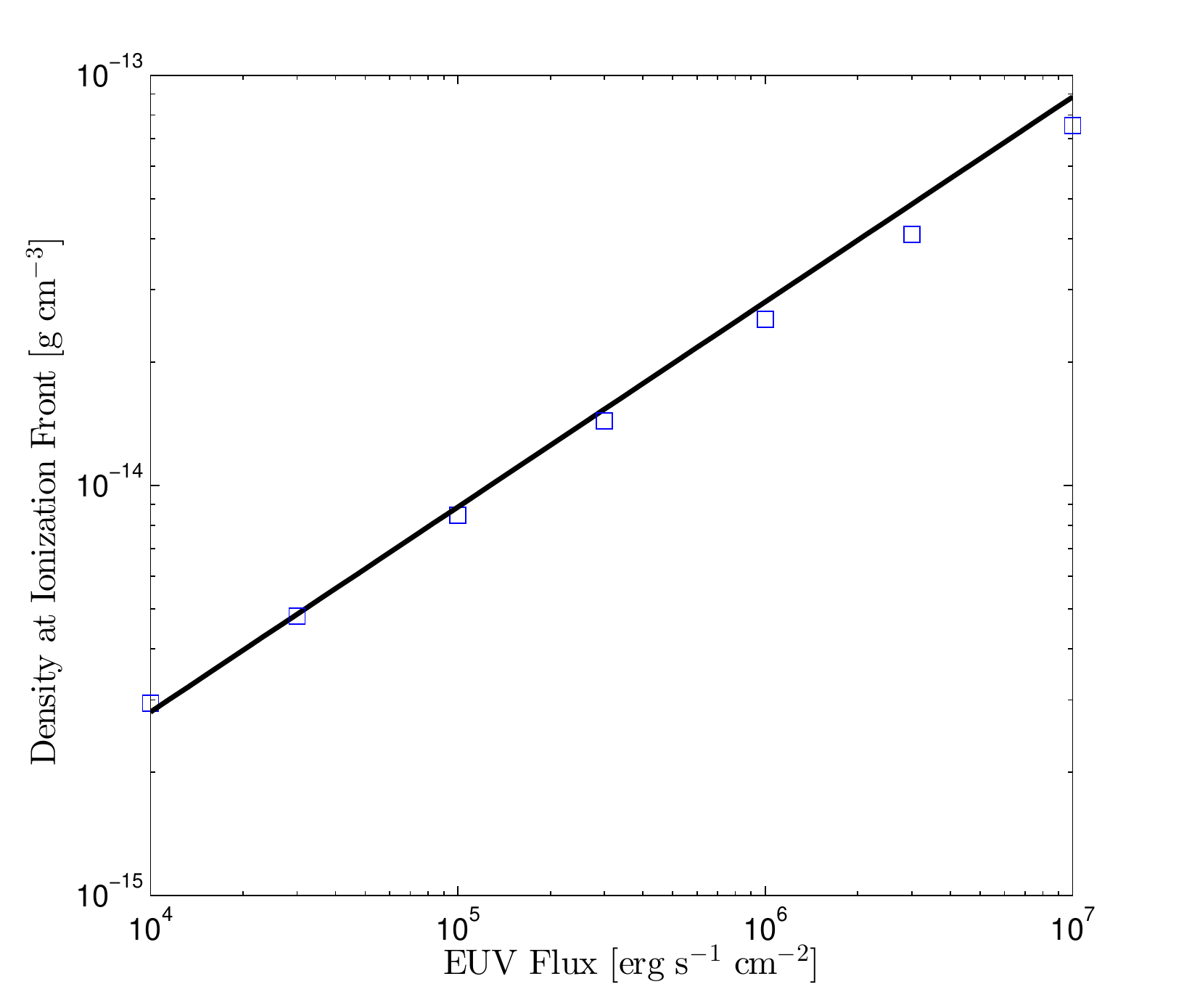}
\caption{Figure showing the results of the ionization front test. 
The points show the simulated values (defined as the density at the
location where the ionization fraction $X=0.9$) and the solid line
shows the analytic expectation given by
equation~(\ref{eqn:density_if}).}
\label{fig:density_test}
\end{figure}

Secondly, in the recombination limit it is well known the density at
the base of the ionization front should scale with the incoming flux
as $n \propto F_{\rm UV}^{1/2}$ \citep[e.g.][]{spitzer}. Furthermore,
assuming the structure to be hydrostatic (true for flows that are
highly sub-sonic near the ionization front), one can calculate the
density at the ionization front by assuming steady state in
equation~(\ref{eqn:ionization}). This is the expresion evaluated in
Section~\ref{sec:dipole} and presented in equation~(\ref{eqn:density_if}). 
In Figure~\ref{fig:density_test} we show the the density at the
ionization front (defined as $X=0.9$) determined by the code compared
to the analytical expectation.

We find that our radiative-transfer and ionization balance algorithm
to be in-agreement with analytic expectations giving us confidence it
is suitable for our evaporation study presented. Finally, we perform
several convergence tests: 1) we double the spatial resolution, 2) we
double the size of our `blocks' in the hybrid characteristics
ray-tracing method detailed in Appendix~\ref{sec:radtransfer}, 3) with
half the size of our ray-tracing blocks. All these tests where
preformed on our $B_P=1$ gauss, $\bratio=0$ and $F_{\rm UV}=10^{6}$
erg~s$^{-1}$~cm$^{-2}$ calculation and show agreement to $\lesssim 5\%$, 
indicating our chosen resolution and `block' size is appropriate for
our calculations.

%\newpage


\begin{thebibliography}{}

\bibitem[Adams(2011)]{adams2011} Adams, F. C. 2011, ApJ, 730, 27 (Paper I) 

\bibitem[Adams \& Gregory(2012)]{ag2012} Adams, F. C., \& Gregory,
  S. G. 2012, ApJ, 744, 55

\bibitem[Adams et al.(2004)]{adams2004} Adams, F. C., Hollenbach, D.,
  Laughlin, G., \& Gorti, U. 2004, ApJ, 611, 360

\bibitem[Baliunas et al.(1996)]{bali} Baliunas, S., Sokoloff, D., \&
  Soon, W. 1996, ApJ, 457, L99

\bibitem[Banaszkiewicz et al.(1998)]{banz} Banaszkiewicz, M., Axford,
  W. I., \& McKenzie, J. F. 1998, A\&A, 337, 940

\bibitem[Baraffe et al.(2006)]{barf1} Baraffe, I., Alibert, Y.,
  Chabrier, G., \& Benz, W. 2006, A\&A, 450, 1221

\bibitem[Baraffe et al.(2004)]{barf2} Baraffe, I., Selsis, F.,
  Chabrier, G., Barman, T. S., Allard, F., Hauschildt, P. H., \&
  Lammer, H. 2004, A\&A, 419, L13

\bibitem[Black(1981)]{black} Black, J. H. 1981, MNRAS, 197, 553 

\bibitem[Blandford \& Payne(1982)]{blan82} Blandford, R.~D., \& Payne,
  D.~G. 1982, MNRAS, 199, 883

\bibitem[Bodenheimer et al.(2003)]{boden2003} 
Bodenheimer, P., Laughlin, G., \& Lin, D.N.C. 2003, ApJ, 592, 555  

\bibitem[Boue et al.(2012)]{boue} Bou{\'e}, G., Figueira, P., Correia,
  A.C.M., \& Santos, N. C. 2012, A\&A, 537, L3 
%Orbital migration induced by anisotropic evaporation. 
%Can hot Jupiters form hot Neptunes?

\bibitem[Cohen et al.(2009)]{cohen09} Cohen, O., Drake, J. J.,
  Kashyap, V. L., Saar, S. H., Sokolov, I. V., Manchester, W. B.,
  Hansen, K. C., \& Gombosi, T. I. 2009, ApJ, 704, 85 
%Interactions of the Magnetospheres of Stars and Close-In Giant Planets

\bibitem[Cuntz et al.(2000)]{cuntz} Cuntz, M., Saar, S. H., \&
  Musielak, Z. E. 2000, ApJ, 533, 151

\bibitem[Donati et al.(1997)]{donati97} Donati, J.-F., Semel, M.,
  Carter, B.~D., Rees, D.~E., \& Collier Cameron, A. 1997, MNRAS,
  291, 658
  
\bibitem[Erkaev et 
al.(2007)]{erkaev07} Erkaev, N.~V., Kulikov, Y.~N., Lammer, H., et al.\ 2007, A\&Ap, 472, 329 
  
\bibitem[Evans 
\& Hawley(1988)]{ct} Evans, C.~R., \& Hawley, J.~F.\ 1988, ApJ, 332, 659 

\bibitem[Garcia-Munoz(2007)]{garcia} Garc{\'i}a Mu{\~n}oz, A. 2007,
  Planet. Space Sci., 55, 1426

\bibitem[Ghosh \& Lamb(1978)]{gho78} Ghosh, P., \& Lamb, F.~K. 1978,
  ApJ, 223, L83

\bibitem[Ghosh \& Lamb(1979)]{ghosh} Ghosh, P., \& Lamb, F. K. 1979,
  ApJ, 232, 259

\bibitem[Gregory et al.(2006)]{gregory2006} Gregory, S.~G., Jardine, M.,
  Simpson, I., \& Donati, J.-F. 2006, MNRAS, 371, 999

\bibitem[Gregory et al.(2008)]{gregory2008} Gregory, S.~G., Matt, S.~P.,
  Donati, J.-F., \& Jardine, M. 2008, MNRAS, 389, 1839

\bibitem[Gregory et al.(2010)]{gregory2010} Gregory, S.~G., Jardine, M.,
  Gray, C.~G., \& Donati, J.-F. 2010, Reports on Progress in Physics,
  73, 126901

\bibitem[Gregory(2011)]{gregory2011} Gregory, S. G. 2011,
  Am. J. Phys., 79, 461

\bibitem[Griesmeier et al.(2004)]{gries} Griesmeier, J.-M.,
  Stadelmann, A., Penz, T., Lammer, H., Selsis, F., Ribas, I., Guinan,
  E. F., Motschmann, U., Biernat, H. K.m \& Weiss, W. W. 2004, A\&A,
  425, 753

\bibitem[Gritschneder et al.(2009)]{ivine} Gritschneder, M., 
Naab, T., Burkert, A., et al.\ 2009, MNRAS, 393, 21


\bibitem[Guenther \& Emerson(1997)]{gunemer} Guenther, E. W., \&
  Emerson, J. P. 1997, A\&A, 321, 803

\bibitem[Hartigan et al.(1995)]{hart} Hartigan, P., Edwards, S., \&
  Ghandour, L. 1995, ApJ, 452, 736

\bibitem[Hayes et al.(2006)]{hayes06} Hayes, J.~C., Norman, 
M.~L., Fiedler, R.~A., et al.\ 2006, ApJS, 165, 188 

\bibitem[Ip et al.(2004)]{ip} Ip, W.-H., Kopp, A., \& Hu, J.-H. 2004,
  ApJ, 602, L53

\bibitem[Johns-Krull(2009)]{jk2009} Johns-Krull, C. M. 2009, IAU
  Symposium, 259, 345

\bibitem[Khodachenko et al.(2012)]{khod} Khodachenko, M. L., Alexeev,
  I., Belenkaya, E., Lammer, H., Grießmeier, J.-M., Leitzinger, M.,
  Odert, P., Zaqarashvili, T., Rucker, H. O. 2012, ApJ, 744, 70
%Magnetospheres of Hot Jupiters: 
%The Importance of Magnetodisks in Shaping a Magnetospheric Obstacle

\bibitem[K{\"o}nigl(1991)]{kon91} K{\"o}nigl, A. 1991, ApJ, 370, L39 

\bibitem[Koskinen et al.(2007)]{koskinen07} Koskinen, T.~T., 
Aylward, A.~D., \& Miller, S.\ 2007, Nature, 450, 845 

\bibitem[Koskinen et al.(2010)]{koskinen10} Koskinen, T.~T., Cho, 
J.~Y.-K., Achilleos, N., \& Aylward, A.~D.\ 2010, ApJ, 722, 178


\bibitem[Koskinen et al.(2013)]{koskinen13} Koskinen, T.~T., 
Harris, M.~J., Yelle, R.~V., \& Lavvas, P.\ 2013, Icaurus, 226, 1678 

\bibitem[Laine et al.(2008)]{laine2008} Laine, R. O., Lin, D.N.C., \&
  Dong, S. 2008, ApJ, 685, 521

\bibitem[Lammer et al.(2003)]{lammer} Lammer, H., Selsis, F., Ribas,
  I., Guinan, E. F., Bauer, S. J., \& Weiss, W. W. 2003, ApJ, 598, 121

\bibitem[Lanza(2008)]{lanza08} Lanza, A. F. 2008, A\&A, 487, 1163 

\bibitem[Lanza(2009)]{lanza09} Lanza, A. F. 2009, A\&A, 505, 339 
% energy production/dissipation via planet forcing reconnection 
% field lines are connected to the star, leading to a chromospheric 
% hot spot rotating synchronously with the planet. 

\bibitem[Lanza(2012)]{lanza12} Lanza, A. F. 2012, A\&A, 544, 23
% energy production/dissipation via planet forcing reconnection

\bibitem[Laughlin et al.(2011)]{laugh2011} 
Laughlin, G., Crismani, M., \& Adams, F. C. 2011, ApJ, 729, 7L 

\bibitem[Lecavelier des Etangs et al.(2010)]{lev10} Lecavelier des
  Etangs, A., Ehrenreich, D., Vidal-Madjar, A., Ballester, G. E.,
  D\'esert, J.-M., Ferlet, R., H\'ebrard, G., Sing, D. K.,
  Tchakoumegni, K.-O., \& Udry, S. 2010, A\&A, 514, 72 
% Evaporation of the planet HD189733b observed in HI Lyman-alpha

\bibitem[Linsky et al.(2010)]{linsky} Linsky, J. L., Yang, H., France,
  K., Froning, C. S., Green, J. C., Stocke, J. T., \& Osterman, S. N.
  2010, ApJ, 717, 1291


\bibitem[Mellema et al.(2006)]{mellema06} Mellema, G., Iliev, 
I.~T., Alvarez, M.~A., \& Shapiro, P.~R.\ 2006, New Astronomy, 11, 374 

\bibitem[Murray-Clay et al.(2009)]{mc2009} Murray-Clay, R. A., Chiang,
  E. I., \& Murray, N. 2009, ApJ, 693, 23

\bibitem[Osterbrock(1989)]{osterbrock} Osterbrock, D.~E.\ 1989, 
Research supported by the University of California, John Simon Guggenheim 
Memorial Foundation, University of Minnesota, et al.~Mill Valley, CA, 
University Science Books, 1989, 422 p.,

\bibitem[Owen et al.(2010)]{owen2010} Owen, J. E., Ercolano, B.,
  Clarke, C. J., \& Alexander, R. D. 2010, MNRAS, 401, 1415
  
\bibitem[Owen et al.(2012)]{owen12} Owen, J.~E., Clarke, 
C.~J., \& Ercolano, B.\ 2012, MNRAS, 422, 1880 

\bibitem[Owen 
\& Jackson(2012)]{oj12} Owen, J.~E., \& Jackson, A.~P.\ 2012, MNRAS, 425, 2931

\bibitem[Owen \& Wu(2013)]{owenwu} Owen, J. E., \& Wu, Y. 2013, ApJ, 775, 105 

\bibitem[Parker(1958)]{parkerspiral} Parker, E. N. 1958, ApJ, 128, 664

\bibitem[Parker(1965)]{park65} Parker, E.~N. 1965, Space Science Reviews, 4, 666

\bibitem[Preusse et al.(2005)]{preusse} Preusse, S., Kopp, A.,
  B{\"u}chner, J., \& Motschmann, U. 2004, A\&A, 434, 1191

\bibitem[Radoski(1967)]{radoski} Radoski, H. R. 1967, JGR, 72, 418

\bibitem[Rijkhorst et al.(2006)]{hybrid_char} Rijkhorst, E.-J., Plewa,
  T., Dubey, A., \& Mellema, G.\ 2006, A\&Ap, 452, 907

\bibitem[Romanova et al.(2002)]{rom02} Romanova, M.~M., Ustyugova,
  G.~V., Koldoba, A.~V., \& Lovelace, R.V.E. 2002, ApJ, 578, 420 

\bibitem[Romanova et al.(2003)]{rom03} Romanova, M.~M., Ustyugova,
  G.~V., Koldoba, A.~V., Wick, J.~V., \& Lovelace, R.V.E. 2003,
  ApJ, 595, 1009

\bibitem[Salat \& Tataronis(2000)]{salat} Salat, A., \& Tataronis,
  J. A. 2000, J. Geophys. Res., 105, 13055 

\bibitem[Schmidt-Voigt \& Koeppen(1987)]{SchmidtVoigt87}
  Schmidt-Voigt, M., \& Koeppen, J.\ 1987, A\&Ap, 174, 211

\bibitem[Shkolnik et al.(2005)]{shkolnik2005} Shkolnik, E., Walker,
  G.A.H., Bohlender, D. A., Gu, P.-G., \& K{\"u}rster, M. 2005, ApJ,
  622, 1075

\bibitem[Shkolnik et al.(2008)]{shkolnik2008} Shkolnik, E., Bohlender,
  D. A., Walker, G.A.H., Collier Cameron, A.  2008, ApJ, 676, 628

\bibitem[Shu(1992)]{shu92} Shu, F. H. 1992, Gas Dynamics (Mill Valley:
  Univ. Science Books)

\bibitem[Skumanich(1972)]{skum} Skumanich, A. P. 1972, ApJ, 171, 565

\bibitem[Spitzer(1978)]{spitzer} Spitzer, L. 1978, Physics Processes
  in the Interstellar Medium (New York: Wiley)

\bibitem[Stone \& Norman(1992a)]{stone_hd} Stone, J.~M., \& Norman,
  M.~L.\ 1992, ApJS, 80, 753

\bibitem[Stone \& Norman(1992b)]{stone_mhd} Stone, J.~M., \& Norman,
  M.~L.\ 1992, ApJS, 80, 791

\bibitem[Stone \& Proga(2009)]{stone} Stone, J. M., \& Proga, D. 2009,
  ApJ, 694, 205

\bibitem[Trammell et al.(2011)]{trammell2011} Trammell, G. B., Arras, P.,
  \& Li, Z.-Y. 2011, ApJ, 728, 152 

\bibitem[Trammell et al.(2014)]{trammell2014} Trammell, G. B., Arras, P.,
  \& Li, Z.-Y. 2014, ApJ, in press

\bibitem[Vidal-Madjar et al.(2003)]{vidal} Vidal-Madjar, A.,
  Lecavelier des Etangs, A., D{\'e}sert, J.-M., Ballester, G. E.,
  Ferlet, R., H{\'e}brard, G., \& Mayor, M. 2003, Nature, 422, 143

\bibitem[Waston et al.(1981)]{watson} Watson, A., Donahue, T., \&
  Walker, J. 1981, Icarus, 48, 150

\bibitem[Weber \& Davis(1967)]{weberdavis} Weber, E. J., \& Davis,
  L. 1967, ApJ, 148, 217 

\bibitem[Weinreich(1998)]{wein} Weinreich, G. 1998, Geometrical
  Vectors (Chicago: Univ. Chicago Press)

\bibitem[Wiktorowicz(2009)]{sloane} Wiktorowicz, S. J. 2009, ApJ,
  696, 1116

\bibitem[Wise \& Abel(2011)]{moray} Wise, J.~H., \& Abel, T.\ 2011,
  MNRAS, 414, 3458

\bibitem[Wood et al.(2002)]{wood} Wood, B., M{\" u}ller, H.-R., Zank,
  G., \& Linsky, J., 2002, ApJ, 574, 412 
%Mass-Loss Rates of Solar-like Stars vs Age and Activity

\bibitem[Woods et al.(1998)]{woods} Woods, T. N., Rottman, G. J.,
  Bailey, S. M., Solomon, S. C., \& Worden, J. R.  1998, Sol. Phys.,
  177, 133

\bibitem[Wright(1959)]{wright} Wright, E. M. 1959,
  Bull. Amer. Math. Soc. 65, 89

\bibitem[Yelle(2004)]{yelle} Yelle, R. V. 2004, Icarus, 170, 167


\end{thebibliography}
\end{document}